\documentclass[aps,prd,12pt,notitlepage,showpacs,nofootinbib,tightenlines]{revtex4-1}
\usepackage{amsmath}
\usepackage{amssymb}
\usepackage{epsfig}
\usepackage{graphicx}
\usepackage{bm}
\usepackage{times}
\usepackage{braket}
\usepackage{color}
\usepackage{slashed}
\usepackage{hyperref}
\definecolor{Red}{rgb}{1.,0.,0.}

\definecolor{Blue}{rgb}{0.,0.,1.}

\definecolor{nicered}{rgb}{0.7,0.1,0.1}
\definecolor{nicegreen}{rgb}{0.1,0.5,0.1}
\hypersetup{colorlinks,citecolor=nicegreen,linkcolor=nicered}

\begin{document}

\newcommand{\beq}{\begin{eqnarray}}
\newcommand{\eeq}{\end{eqnarray}}
\newcommand{\non}{\nonumber\\ }

\newcommand{\jpsi}{J/\Psi}

\newcommand{\ppa}{\phi_\pi^{\rm A}}
\newcommand{\ppp}{\phi_\pi^{\rm P}}
\newcommand{\ppt}{\phi_\pi^{\rm T}}
\newcommand{\ov}{ \overline }

\newcommand{\zerot}{ {\textbf 0_{\rm T}} }
\newcommand{\kt}{k_{\rm T} }
\newcommand{\fb}{f_{\rm B} }
\newcommand{\fk}{f_{\rm K} }
\newcommand{\rk}{r_{\rm K} }
\newcommand{\mb}{m_{\rm B} }
\newcommand{\mw}{m_{\rm W} }
\newcommand{\im}{{\rm Im} }

\newcommand{\kks}{K^{(*)}}
\newcommand{\acp}{{\cal A}_{\rm CP}}
\newcommand{\pb}{\phi_{\rm B}}

\newcommand{\xeba}{\bar{x}_2}
\newcommand{\xsba}{\bar{x}_3}
\newcommand{\peas}{\phi^A}

\newcommand{\pvsl}{ p \hspace{-2.0truemm}/_{K^*} }
\newcommand{\esl}{ \epsilon \hspace{-2.1truemm}/ }
\newcommand{\psl}{ p \hspace{-2truemm}/ }
\newcommand{\ksl}{ k \hspace{-2.2truemm}/ }
\newcommand{\lsl}{ l \hspace{-2.2truemm}/ }
\newcommand{\nsl}{ n \hspace{-2.2truemm}/ }
\newcommand{\vsl}{ v \hspace{-2.2truemm}/ }
\newcommand{\epsl}{\epsilon \hspace{-1.8truemm}/\,  }
\newcommand{\bfkk}{{\bf k} }
\newcommand{\calm}{ {\cal M} }
\newcommand{\calh}{ {\cal H} }
\newcommand{\calo}{ {\cal O} }

\def \appb{{\bf Acta. Phys. Polon. B }  }
\def \cpc{ {\bf Chin. Phys. C } }
\def \ctp{ {\bf Commun. Theor. Phys. } }
\def \epjc{{\bf Eur. Phys. J. C} }
\def \jhep{{\bf J. High Energy Phys. } }
\def \jpg{ {\bf J. Phys. G} }
\def \mpla{{\bf Mod. Phys. Lett. A } }
\def \npb{ {\bf Nucl. Phys. B} }
\def \plb{ {\bf Phys. Lett. B} }
\def \pr{  {\bf Phys. Rep.} }
\def \prc{ {\bf Phys. Rev. C }}
\def \prd{ {\bf Phys. Rev. D} }
\def \prl{ {\bf Phys. Rev. Lett.}  }
\def \ptp{ {\bf Prog. Theor. Phys. }}
\def \zpc{ {\bf Z. Phys. C}  }
\def \jpg{ {\bf J.Phys.-G-}  }
\def \ap{ {\bf Ann. of Phys}  }

\title{The perturbative QCD factorization of $\rho \gamma^{\star} \to \pi$}
\author{Shan Cheng$^{1}$} \email{chengshan-anhui@163.com}
\author{Zhen-Jun Xiao$^{1,2}$ } \email{xiaozhenjun@njnu.edu.cn}
\affiliation{1.  Department of Physics and Institute of Theoretical Physics,
Nanjing Normal University, Nanjing, Jiangsu 210023, People's Republic of China,}
\affiliation{2. Jiangsu Key Laboratory for Numerical Simulation of Large Scale Complex Systems,
Nanjing Normal University, Nanjing 210023, People's Republic of China}
\date{\today}
\vspace{1cm}
\begin{abstract}
In this paper, we firstly varify that the factorization hypothesis is valid for the exclusive process
$\rho \gamma^{\star} \to \pi$ at the next-to-leading order (NLO) with the collinear
factorization approach, and then extend this proof to the case of the
$k_T$ factorization approach.
We particularly show that at the NLO level, the soft divergences in the full
quark level calculation could be canceled completely as for the $\pi \gamma^{\star} \to \pi$
process where only the pseudoscalar $\pi$ meson involved,
and the remaining collinear divergences can be absorbed into the NLO hadron wave functions.
The full amplitudes can be factorized as the convolution of the NLO wave functions
and the infrared-finite hard kernels with these factorization approaches.
We also write out the NLO meson distribution amplitudes in the form of
nonlocal matrix elements.
\end{abstract}



\maketitle

\section{Introduction}

As the fundamental tool of the perturbative Quantum Chromodynamics(QCD)\cite{Collins pqcd}
with a large momentum translation,
the factorization theorem \cite{plb87-359} assume that the hard part of the relevant  processes
is infrared-finite and can be calculated,
while the non-perturbative dynamics of these high-energy QCD processes can be canceled
at the quark level or absorbed into the input universal hadron wave functions.
The physical quantities can be written as the convolutions of the hard part kernels and the universal
processes-independent wave functions, and then the perturbative QCD has the prediction power.
The collinear factorization \cite{Sterman,prl83-1914} and the $k_T$ factorization
\cite{npb360-3,zpz50-139,npb366-135},
with the distinction whether to keep the transversal momenta in the propergators,
are the two popular factorization approaches applied on the kard QCD processes.

We know that the  theoretical study  for the exclusive processes are in general
more difficult than that for the inclusive processes \cite{npb529-323}.
Because in the exclusive processes, the pQCD factorization in it's standard form may
be valid only for the large momentum transfer processes;
while in the inclusive processes, like the deep-inelastic scattering,
the leading twist factorization approximation is adequate already at $Q \sim 1$ Gev.
So the intensively investigation for the factorization theorems or  the factorization approaches
for the exclusive processes is unavoidable.

In recent years, based on the factorization hypothesis, the collinear factorization and $k_T$ factorization
for the exclusive processes $\pi \gamma^{\star} \to \gamma(\pi)$
and $B \to \gamma(\pi) l \overline{\nu}$
have been testified both at the leading order (LO) and the next-to-leading order (NLO)
level, and then these factorization proofs were developed into all-orders with the induction
approach\cite{prd64-014019,prd67-034001,epjc40-395}.
The NLO hard kernels for these exclusive processes
have also been calculated for example in
Refs.~\cite{prd76-034008,prd83-054029,prd85-074004,prd89-054015,prd89-094004}.
These NLO evaluations showed that the positive corrections from the leading twist
would be cancelled partly by the negative corrections from the NLO twist,
resulting in a small net NLO correction to the leading order hard kernels,
which further verified the feasibility of the perturbative QCD to those considered
exclusive processes.
But all these proofs and calculations are only relevant for the pseudo-scalar mesons,
the exclusive processes with vector mesons have not been included at present.
The study of the electromagnetic form factor processes between the vector meson and the
pseudo-scalar meson is an important way to understand the internal structure of hadrons.
There are many works  on this subject:
(a) $\rho$ meson transition and electromagnetic form factors are predicted at the
NLO level in the QCD sum rule analysis\cite{prd70-033001};
(b) space-like and time-like pion-rho transition form factors were investigated in
Ref.~\cite{jpg34-1845} in the light-cone formalism;
(c) the meson transition form factors were studied within a model of QCD based
on the Dyson-Schwinger equations in\cite{prc65-045211};
and (d) the transition form factor of $\rho \gamma^{\star} \to \pi$ was also extracted
from the other processes in the extended hard-wall AdS/QCD model\cite{epjc67-253} recently.

In this paper, we also consider  the rho-pion transition process.
By inserting the Fierz identity into the relevant expressions and employing the eikonal approximation,
we can factorize the fermion flow and the momentum flow effectively.
By summing over all the color factors, we can express these irreducible convolutions
into three parts: with the additional gluon momentum flow, not flow and partly flow into
the leading order hard kernel.
We will do the factorization proof for the exclusive process $\rho \gamma^{\star} \to \pi$
at the NLO level,
from the collinear factorization to the $k_T$ factorization approach.
With the light-cone kinetics, we will obtain the gauge invariant nonlocal matrix element
for the pion meson and rho meson
wave functions along the light-cone direction in the collinear factorization,
and lightly deviate from the light-cone direction in the $k_T$ factorization.
At the NLO level, we clearly verified that the soft divergences will be canceled in
the quark level diagrams, and the collinear divergences can be absorbed into the NLO wave functions,
then we can obtain an infrared-finite next-to-leading order hard kernel in principle.

The paper is organized as following. The leading order dynamical analysis is presented in the second section.
In section-III we prove that the collinear factorization approach is valid for the $\rho \to \pi$ transition
process at the next-to-leading order.
The collinear factorization approach is extended to the $k_T$ factorization approach
for this $\rho \to \pi$ transition process in section-VI.
The summary and some discussions will appear at the final section.


\section{Collinear Factorization Of $\rho \gamma^{\star} \to \pi$}

In this section we will prove  the collinear factorization of the transition
$\rho \gamma^{\star} \to \pi$.
We firstly consider the two sets of leading order transition amplitudes, and then use
the Fierz identity and the eikonal
approximation to factorize the fermion currents and the momentum currents
at the NLO level, in order to obtain the NLO transition amplitudes for each sub-diagram
in the convoluted forms of the LO hard transition amplitudes and the gauge
invariant nonlocal NLO distribution
amplitudes(DAs) along the light-core(LC) direction.
We finally sum up all the sub-diagrams for each set  to collect all the color factors.
The key point of the factorization is to find and absorb the infrared divergences,
so we will not consider the self-energy corrections to the internal quark lines
because they don't generate infrared divergences.

\subsection{Leading Order Hard Kernel}
\begin{figure}[tb]
\vspace{-1cm}
\begin{center}
\leftline{\epsfxsize=12cm\epsffile{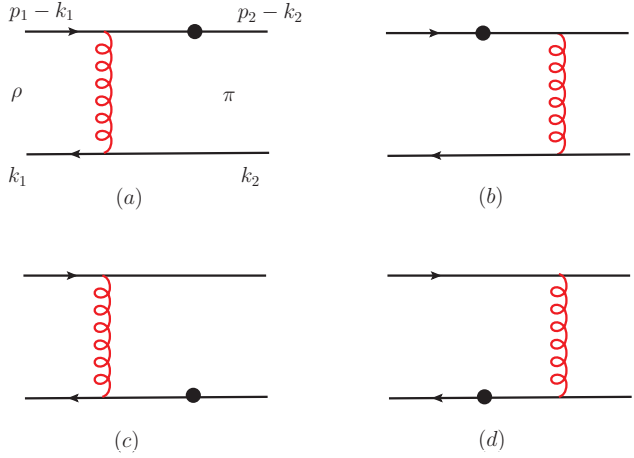}}
\end{center}
\vspace{-10cm}
\caption{The four leading-order quark diagrams for the $\rho \gamma^{\star} \to \pi$
form factor with the symbol $\bullet$ representing the virtual photon vertex.}
\label{fig:fig1}
\end{figure}

The LO quark diagrams for the $\rho \gamma^{\star} \to \pi$ transition are shown in the Fig.~\ref{fig:fig1},
where the virtual photon vertex represented by the dark spot have been placed at the four different
positions respectively.
In the light-cone coordinator system, the incoming $\rho$ meson carry the momenta
 $p_1 = \frac{Q}{\sqrt{2}}(1,0,\mathbf{0_T})$,
and the outgoing $\pi$ carry the momenta $p_2 = \frac{Q}{\sqrt{2}}(0,1,\mathbf{0_T})$.
Besides the momenta, the initial $\rho$ would carry the longitudinal polarization vector
$\epsilon_{1 \mu}(L) = \frac{1}{\sqrt{2} \gamma_{\rho}}(1,-\gamma_{\rho},\mathbf{0_{T}})$ and the
transversal polarization vector $\epsilon_{1 \mu}(T) = (0,0,\mathbf{1_{T}})$.
The momenta carried by the anti-quark of the initial and final state meson are defined as
$k_1=\frac{Q}{\sqrt{2}}(x_1,0,\mathbf{0_T})$ and $k_2 = \frac{Q}{\sqrt{2}}(0,x_2,\mathbf{0_T})$
with $x_1$ and $x_2$ being the momentum fraction carried by the anti-partons inside $\rho$ and $\pi$.

As the spin-1 particle, the wave functions for $\rho$ meson should contain both longitudinal and
transverse components\cite{prd65014007}.
\beq
\Phi_{\rho}(p_1,\epsilon_{1T})&=&\frac{i}{\sqrt{2N_{c}}}
\left [M_{\rho} \esl_{1T} \phi^{v}_{\rho}(x_{1})+\esl_{1T} \psl_{1}\phi^{T}_{\rho}(x_{1})
+M_{\rho} i \epsilon_{\mu' \nu \rho \sigma} \gamma_{5} \gamma^{\mu'} \epsilon^{\nu}_{1T} n^{\rho} v^{\sigma} \phi^{a}_{\rho}(x_{1})\right ],\non
\Phi_{\rho}(p_1,\epsilon_{1L})&=&\frac{i}{\sqrt{2N_{c}}}\left [M_{\rho} \esl_{1L} \phi_{\rho}(x_{1})
+\esl_{1L} \psl_{1}\phi^{t}_{\rho}(x_{1}) + M_{\rho} \phi^{s}_{\rho}(x_{1})\right ],
\label{eq:wfrho}
\eeq
in which $\phi_{\rho}$ and $\phi^{T}_{\rho}$ are twist-2 (T2) DAs,
$\phi^{t/s}_{\rho}, \phi^{t/s}_{\rho}$ are twist-3 (T3) DAs,
and the unit vector $n/v$ is defined as $(1,0,\mathbf{0})/(0,1,\mathbf{0})$.
The pseudoscalar $\pi$ meson wave function up to twist-3 is also given as
in Refs.~\cite{prd71014015,jhep0605004,prd90014029}
\beq
\Phi_{\pi}(p_2)=\frac{-i}{\sqrt{2N_{c}}}\left \{\gamma_5 \psl_{2} \phi^{a}_{\pi}(x_{2})
+ m_0^{\pi}\gamma_{5}\left [\phi^{p}_{\pi}(x_{2})+(\vsl \nsl-1)\phi^{t}_{\pi}(x_{2})\right ]\right \},
\label{eq:wfpi}
\eeq
with the twist-2 DA $\phi^{a}_{\pi}$ and twist-3 DAs $\phi^{p}_{\pi}$ and $\phi^{t}_{\pi}$.
The operator product expansion (OPE)\cite{ap77-536} states that amplitudes from the twist-3 DAs
are suppressed by the hierarchy $M_{\rho}/Q$ and $m_0^\pi/Q$ at the large momenta transition
region, when compared with the twist-2 DAs of the $\rho$ and $\pi$ meson wave functions respectively.
We can classify the LO transition amplitudes into four sets by the twists' analysis of the
initial and final meson wave functions: T2\&T2; T2\&T3; T3\&T2 and finally T3\&T3.
Fortunately, we just need to consider the first two sub-diagrams
Fig.~\ref{fig:fig1}(a) and Fig.~\ref{fig:fig1}(b) directly,
because the amplitudes of sub-diagram Fig.~\ref{fig:fig1}(c) (Fig.~\ref{fig:fig1}(d) )
can be obtained by simple replacement $x_i \to 1-x_i (i=1,2)$ from the amplitudes of
Fig.~\ref{fig:fig1}(a) ( Fig.~\ref{fig:fig1}(b)).
The standard calculations show that only the T3\&T2 set (  the twist-3 DAs of the rho meson
and the twist-2 DAs of the pion meson)  contribute to the LO transition amplitude
of Fig.~\ref{fig:fig1}(a), which can be written as the following form,
\beq
G^{(0)}_{a,32}(x_1,x_2) = \frac{i e g^2_s C_F}{2} \frac{[\epsl_{1T} M_{\rho} \phi^{v}_{\rho}
+ M_{\rho} i \epsilon_{\mu'\nu\rho\sigma} \gamma_5 \gamma^{\mu'} \epsilon^{\nu}_{1T} n^{\rho} v^{\sigma} \phi^{a}_{\rho}]
\gamma^{\alpha} [\gamma_5 \psl_2 \phi^A_{\pi}] \gamma_{\mu} (\psl_1 - \ksl_2) \gamma_{\alpha}}{(p_1-k_2)^2 (k_1-k_2)^2},
\label{eq:loa32}
\eeq
where $\gamma^{\alpha}$ should be chosen as $\gamma^-$.
Similarly, only the crossed sets of T2\&T3 (Set-I) and T3\&T2 (Set-II) contribute to the LO transition
amplitudes of Fig.~\ref{fig:fig1}(b), which can be written as the form of
\beq
G^{(0)}_{b,23}(x_1,x_2) = \frac{i e g^2_s C_F}{2} \frac{[\epsl_{1T} \psl_1 \phi^{T}_{\rho}]
\gamma^{\alpha} [\gamma_5 m^0_{\pi} \phi^P_{\pi}] \gamma_{\alpha} (\psl_2 - \ksl_1) \gamma_{\mu}}{(p_2-k_1)^2 (k_1-k_2)^2},
\label{eq:lob23}
\eeq
where the $\gamma^{\alpha}$ can be $\gamma^-$ or $\gamma^{\alpha}_{\bot}$;
\beq
G^{(0)}_{b,32}(x_1,x_2) = \frac{i e g^2_s C_F}{2} \frac{[\epsl_{1T} M_{\rho} \phi^{v}_{\rho}
+ M_{\rho} i \epsilon_{\mu'\nu\rho\sigma} \gamma_5 \gamma^{\mu'} \epsilon^{\nu}_{1T} n^{\rho} v^{\sigma} \phi^{a}_{\rho}]
\gamma^{\alpha} [\gamma_5 \psl_2 \phi^A_{\pi}] \gamma_{\alpha} (\psl_2 - \ksl_1) \gamma_{\mu}}{(p_2-k_1)^2 (k_1-k_2)^2},
\label{eq:lob32}
\eeq
where the $\gamma^{\alpha} = \gamma^{\alpha}_{\bot}$.
The LO transition amplitudes as given in Eqs.~(\ref{eq:loa32},\ref{eq:lob23},\ref{eq:lob32})
are all transversal due to the $\gamma_5$ from the final pion meson wave function, the
$\gamma_\mu$ from the virtual photon vertex and the polarization vector $\epsilon_1$ of the initial
$\rho$ meson.

From the expressions of the LO transition amplitudes $G^{(0)}(x_1,x_2)$ as given in
Eqs.~(\ref{eq:loa32},\ref{eq:lob23},\ref{eq:lob32}), one can see that there are
clear qualitative differences between the $\rho \gamma^\star \to \pi$ studied in this paper and
the $\pi \gamma^\star \to \pi$ investigated previously in Refs.~\cite{prd64-014019,epjc40-395,prd83-054029,prd89-054015}:
\begin{enumerate}
\item[(i)]
In the $\pi \gamma^\star \to \pi$ transition,  the initial and final state meson are the same pion.
Consequently, only the contribution from Fig.~\ref{fig:fig1}(a) should be calculated explicitly, while
the contributions from Figs.~\ref{fig:fig1}(b,c,d) can be obtained from those of Fig.~1(a) by
direct kinetic transformations \cite{prd83-054029,prd89-054015}.
Furthermore, only the T2\&T2 and T3\&T3 terms contribute to the LO transition amplitudes because
of the presence of matrix $\gamma_5$ in both the initial and final state pion meson.

\item[(ii)]
For $\rho \gamma^\star \to \pi$ transition, however, the initial and final state meson are the vector
$\rho$ and pseudo-scalar pion. The possible contributions from Figs.~\ref{fig:fig1}(a) and 1(b)
are rather different and should be calculated explicitly.
For $\rho \gamma^\star \to \pi$ transition, in fact, only the transversal component $\Phi_{\rho}(p_1,\epsilon_{1T})$
of initial rho meson in Eq.~(\ref{eq:wfrho}) contribute to the LO rho-pion transition amplitude,
and this LO transition amplitude receive the contributions from $G^{(0)}_{a,32}(x_1,x_2)$ in Eq.~(\ref{eq:loa32} )(i.e.
the crossed-set T3\&T2 ) from Fig.~\ref{fig:fig1}(a), and from $G^{(0)}_{b,23}(x_1,x_2)$ and $G^{(0)}_{b,32}(x_1,x_2)$
in Eqs.~(\ref{eq:lob23},\ref{eq:lob32})(i.e. the crossed sets T2\&T3 and T3\&T2 ) from  Fig.~\ref{fig:fig1}(b).

\end{enumerate}

\subsection{${\calo}(\alpha_s)$ corrections to Fig.1(a)}

A complete amplitude for a physical process in QCD is usually defined in three spaces:
the spin space, the momenta space and the color space.
So the factorization theorems need to deal with all these three spaces in the QCD processes.
We can factorize the fermion currents in the spin space by using the Fierz identity,
\beq
I_{ij}I_{lk} &&= \frac{1}{4}I_{ik}I_{lj} + \frac{1}{4}(\gamma_{5})_{ik}(\gamma_{5})_{lj}
               + \frac{1}{4}(\gamma^{\alpha})_{ik}(\gamma^{\alpha})_{lj} \non
             &&+ \frac{1}{4}(\gamma_{5}\gamma^{\alpha})_{ik}(\gamma_{\alpha}\gamma_{5})_{lj}
               + \frac{1}{8}(\sigma^{\alpha\beta}\gamma_{5})_{ik}(\sigma_{\alpha\beta}\gamma_{5})_{lj},
\label{eq:fierz}
\eeq
where I is the identity matrix and $\sigma^{\alpha \beta}$ is defined by
$\sigma^{\alpha \beta}=i[ \gamma^{\alpha}, \gamma^{\beta}]/2$, the different terms in
Eq.~(\ref{eq:fierz}) stand for different twists' contributions.
The eikonal approximation is used to factorize the momenta currents in the momentum space.
And at last we need to sum over all the color factors to obtain the gauge-independent high order DAs.
In this section we will show the NLO factorization of the $\rho \to \pi$ transition process,
according to the LO transition amplitudes expressed
in Eqs.~(\ref{eq:loa32},\ref{eq:lob23},\ref{eq:lob32}) for the sub-diagrams
Figs.~\ref{fig:fig1}(a,b).
We try to factorize these NLO transition amplitudes into the convolutions of the LO hard
amplitudes and the NLO meson DAs.

We here firstly testify that the collinear factorization is valid at the NLO level
for the Fig.~\ref{fig:fig1}(a), where the LO transition amplitude as given
in Eq.~(\ref{eq:loa32}) contains the T3\&T2 contribution only.
So we just need to consider the twist-3 DAs for the initial $\rho$ meson and the twist-2 DA
for the final state $\pi$ meson in this NLO factorization proofs.

\begin{figure}[tb]
\vspace{0cm}
\begin{center}
\leftline{\epsfxsize=10cm\epsffile{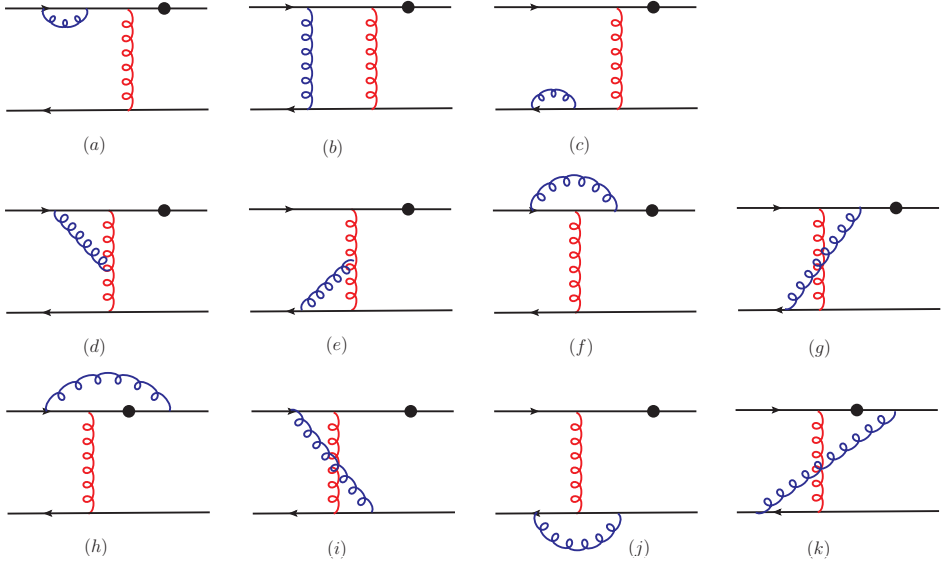}}
\end{center}
\vspace{-6cm}
\caption{$\calo(\alpha_s)$ corrections to Fig.~\ref{fig:fig1}(a) with an additional
gluon (blue curves) emitted from the initial $\rho$ meson.}
\label{fig:fig2}
\end{figure}

There are two types infrared divergences from $\calo(\alpha_s)$ corrections
to Fig.~\ref{fig:fig1}(a) induced by an additional gluon as illustrated in Fig.~\ref{fig:fig2} and Fig.~\ref{fig:fig4},
which are distinguished by the direction of the additional gluon momentum.
We firstly identify these infrared divergences for the $\calo(\alpha_s)$ correction
with the additional "blue" gluon emitted from the initial $\rho$ meson as
shown in Fig.~\ref{fig:fig2}, where the gluon momenta may be parallel to the rho
meson momenta $p_1$.

It's easy to find that the amplitudes in Eqs.~(\ref{eq:2nloa32},\ref{eq:2nlob32},\ref{eq:2nloc32})
are reducible for sub-diagrams Fig.~\ref{fig:fig2}(a,b,c), because we can factorize this
amplitudes by simply inserting the Fierz identity.
The symmetry factor $1/2$ in the self-energy diagrams Eqs.~(\ref{eq:2nloa32},\ref{eq:2nloc32})
represent the freedom to chose the most outside vertex of the additional gluon.
The soft divergences from the $l \sim (\lambda, \lambda, \lambda)$ region are
canceled in these reducible amplitudes
$G^{(1)}_{2a,32}(x_1; x_2), G^{(1)}_{2b,32}(x_1; x_2), G^{(1)}_{2c,32}(x_1; x_2)$,
which is determined by the QCD dynamics that the soft gluon don't resolve the color structure
of the rho meson.

\beq
G^{(1)}_{2a,32} =&& \frac{1}{2} \frac{e g^4_s C^2_F}{2}
    \frac{[\epsl_{1T} M_{\rho} \phi^v_{\rho} + i M_{\rho} \epsilon_{\mu' \nu \rho \sigma} \gamma_5 \gamma^{\mu'}
    \epsilon^{\nu}_{1T} n^{\rho} v^{\sigma} \phi^a_{\rho}]}{(p_1-k_2)^2 (k_1-k_2)^2 (p_1-k_1)^2 (p_1-k_1+l)^2 l^2} \non
 && \cdot \gamma^{\alpha} [\gamma_5 \psl_2 \phi^A_{\pi}] \gamma_{\mu} (\psl_1 - \ksl_2) \gamma_{\alpha} (\psl_1 - \ksl_1)
    \gamma^{\rho'} (\psl_1 - \ksl_1 + \lsl) \gamma_{\rho'} \non
=&& \frac{1}{2} \phi^{(1),v}_{\rho, a} \otimes G^{(0),v}_{a,32}(x_1; x_2)
  + \frac{1}{2} \phi^{(1),a}_{\rho, a} \otimes G^{(0),a}_{a,32}(x_1; x_2),
\label{eq:2nloa32}
\eeq
\beq
G^{(1)}_{2b,32} =&& \frac{- e g^4_s C^2_F}{2}
    \frac{[\epsl_{1T} M_{\rho} \phi^v_{\rho} + i M_{\rho} \epsilon_{\mu' \nu \rho \sigma} \gamma_5 \gamma^{\mu'}
    \epsilon^{\nu}_{1T} n^{\rho} v^{\sigma} \phi^a_{\rho}]}{(p_1-k_2)^2 (k_1-k_2-l)^2 (p_1-k_1+l)^2 (k_1-l)^2 l^2} \non
 && \cdot \gamma^{\rho'} (\ksl_1 - \lsl) \gamma^{\alpha} [\gamma_5 \psl_2 \phi^A_{\pi}] \gamma_{\mu} (\psl_1 - \ksl_2) \gamma_{\alpha}
    (\psl_1 - \ksl_1 + \lsl) \gamma^{\rho'} \non
=&& \phi^{(1),v}_{\rho, b} \otimes G^{(0),v}_{a,32}(\xi_1, x_2)
  + \phi^{(1),a}_{\rho, b} \otimes G^{(0),a}_{a,32}(\xi_1, x_2),
\label{eq:2nlob32}
\eeq
\beq
G^{(1)}_{2c,32} =&& \frac{1}{2} \frac{e g^4_s C^2_F}{2}
    \frac{[\epsl_{1T} M_{\rho} \phi^v_{\rho} + i M_{\rho} \epsilon_{\mu' \nu \rho \sigma} \gamma_5 \gamma^{\mu'}
    \epsilon^{\nu}_{1T} n^{\rho} v^{\sigma} \phi^a_{\rho}]}{(p_1-k_2)^2 (k_1-k_2)^2 (p_1-k_1)^2 (p_1-k_1+l)^2 l^2} \non
 && \cdot \gamma^{\rho'}  (\ksl_1 - \lsl) \gamma_{\rho'} \ksl_1 \gamma^{\alpha} [\gamma_5 \psl_2 \phi^A_{\pi}] \gamma_{\mu}
    (\psl_1 - \ksl_2) \gamma_{\alpha} \non
=&& \frac{1}{2} \phi^{(1),v}_{\rho, c} \otimes G^{(0),v}_{a,32}(x_1, x_2)
  + \frac{1}{2} \phi^{(1),a}_{\rho, c} \otimes G^{(0),a}_{a,32}(x_1, x_2),
\label{eq:2nloc32}
\eeq
where the LO hard amplitudes $G^{(0),v}_{a,32}(\xi_1, x_2)$ and $G^{(0),a}_{a,32}(\xi_1, x_2)$
in Eq.~(\ref{eq:2nlob32})
with the gluon momenta flowing into the LO hard kernel are of the following form
\beq
G^{(0),v}_{a,32}(\xi_1; x_2) = \frac{i e g^2_s C_F}{2} \frac{[\epsl_{1T} M_{\rho} \phi^{v}_{\rho}]
\gamma^{\alpha} [\gamma_5 \psl_2 \phi^A_{\pi}] \gamma_{\mu} (\psl_1 - \ksl_2) \gamma_{\alpha}}{(p_1-k_2)^2 (k_1-k_2-l)^2},
\label{eq:loa32iv}
\eeq
\beq
G^{(0),a}_{a,32}(\xi_1; x_2) = \frac{i e g^2_s C_F}{2}
    \frac{M_{\rho} i \epsilon_{\mu'\nu\rho\sigma} \gamma_5 \gamma^{\mu'} \epsilon^{\nu}_{1T} n^{\rho}v^{\sigma}]
    \gamma^{\alpha} [\gamma_5 \psl_2 \phi^A_{\pi}] \gamma_{\mu} (\psl_1 - \ksl_2) \gamma_{\alpha}}{(p_1-k_2)^2 (k_1-k_2-l)^2}.
\label{eq:loa32ia}
\eeq

The NLO DAs $\phi_\rho^{(1)}$ in Eqs.~(\ref{eq:2nloa32},\ref{eq:2nlob32},\ref{eq:2nloc32}),
which absorbed all the infrared singularities from those reducible sub-diagrams
Figs.~\ref{fig:fig2}(a,b,c), can be  written as the following form
\beq
\phi^{(1),v}_{\rho, a} &=& \frac{- i g^2_s C_F}{4} \frac{\gamma^b_{\bot} \gamma_{\bot b} (\psl_1 - \ksl_1)
    \gamma^{\rho'} (\psl_1 - \ksl_1 + \lsl) \gamma_{\rho'}}{(p_1 -k_1)^2 (p_1 - k_1 + l)^2 l^2}, \non
\phi^{(1),a}_{\rho, a} &=& \frac{- i g^2_s C_F}{4} \frac{\gamma_5 \gamma^{\mu'}_{\bot} \gamma_{\bot \mu'}
    \gamma_5 (\psl_1 - \ksl_1) \gamma^{\rho'} (\psl_1 - \ksl_1 + \lsl) \gamma_{\rho'}}{(p_1 -k_1)^2 (p_1 - k_1 + l)^2 l^2}; \non
\phi^{(1),v}_{\rho, b} &=& \frac{i g^2_s C_F}{4} \frac{\gamma_{\bot}^b \gamma^{\rho'} (\ksl_1 - \lsl)
    \gamma_{\bot b}(\psl_1 - \ksl_1 + \lsl) \gamma_{\rho'}}{(k_1 - l)^2 (p_1 - k_1 + l)^2 l^2}, \non
\phi^{(1),a}_{\rho, b} &=& \frac{i g^2_s C_F}{4} \frac{\gamma_5 \gamma^{\mu'}_{\bot} \gamma^{\rho'}
    (\ksl_1 - \lsl) \gamma_{\bot \mu'} \gamma_5 (\psl_1 - \ksl_1 + \lsl) \gamma_{\rho'}}{(k_1 - l)^2 (p_1 - k_1 + l)^2 l^2};\non
\phi^{(1),v}_{\rho, c} &=& \frac{- i g^2_s C_F}{4} \frac{\gamma^b_{\bot} \gamma^{\rho'}
    (\ksl_1 - \lsl) \gamma_{\rho'} \ksl_1 \gamma_{\bot b}}{(k_1 - l)^2 (k_1 )^2 l^2}, \non
\phi^{(1),a}_{\rho, c} &=& \frac{- i g^2_s C_F}{4} \frac{\gamma_5 \gamma^{\mu'}_{\bot} \gamma^{\rho'}
    (\ksl_1 - \lsl) \gamma_{\rho'} \ksl_1 \gamma_{\bot \mu'} \gamma_5} {(k_1 - l)^2 (k_1 )^2 l^2}.
\label{eq:2nloabcrhova}
\eeq

The additional gluons in sub-diagrams Figs.~\ref{fig:fig2}(d,e,f,g) generate the collinear
divergences only, because one vertex of the gluon is attached to the LO hard part and then the
soft region is strongly suppressed by $1/Q^2$.
For these amplitudes, we choose the radiative gluon momenta being parallel to the
initial rho meson momenta $p_1$ to evaluate the collinear divergences.
All the amplitudes for those sub-diagrams in Fig.~\ref{fig:fig2}(d,e,f,g) are listed in
Eqs.~(\ref{eq:2nlod32},\ref{eq:2nloe32},\ref{eq:2nlof32},\ref{eq:2nlog32}).
For Fig.~\ref{fig:fig2}(d) we find
\beq
G^{(1)}_{2d,32} =&&\frac{- i e g^4_s Tr[T^a T^c T^b] f_{abc}}{2 N_c}
    \frac{[\epsl_{1T} M_{\rho} \phi^v_{\rho} + i M_{\rho} \epsilon_{\mu' \nu \rho \sigma} \gamma_5 \gamma^{\mu'}
    \epsilon^{\nu}_{1T} n^{\rho} v^{\sigma} \phi^a_{\rho}]}{(p_1-k_2)^2 (k_1-k_2)^2 (p_1-k_1+l)^2 (k_1-k_2-l)^2 l^2} \non
 && \cdot \gamma^{\alpha} [\gamma_5 \psl_2 \phi^A_{\pi}] \gamma_{\mu} (\psl_1 - \ksl_2) \gamma^{\beta} (\psl_1 - \ksl_1 + \lsl)
    \gamma^{\gamma} F_{\alpha \beta \gamma} \non
\sim && \frac{9}{16} \phi^{(1),v}_{\rho, d} \otimes [G^{(0),v}_{a,32}(x_1; x_2) - G^{(0),v}_{a,32}(\xi_1; x_2)] \non
    +&& \frac{9}{16} \phi^{(1),a}_{\rho, d} \otimes [G^{(0),a}_{a,32}(x_1; x_2) - G^{(0),a}_{a,32}(\xi_1; x_2)],
\label{eq:2nlod32}
\eeq
with
\beq
\phi^{(1),v}_{\rho, d} &=& \frac{- i g^2_s C_F}{4}
    \frac{\gamma^b_{\bot} \gamma_{\bot b} (\psl_1 - \ksl_1 + \lsl) \gamma^{\rho} v_{\rho}}
    {(p_1 - k_1 + l)^2 l^2 (v \cdot l)}, \non
\phi^{(1),a}_{\rho, d} &=& \frac{- i g^2_s C_F}{4} \frac{(\gamma_5 \gamma^{\mu'}_{\bot})(\gamma_{\bot \mu'} \gamma_5)
    (\psl_1 - \ksl_1 + \lsl) \gamma^{\rho} v_{\rho}}{(p_1 - k_1 + l)^2 l^2 (v \cdot l)}.
\label{eq:2nlodrho}
\eeq
In Eq.~(\ref{eq:2nlod32}), we have
$F_{\alpha \beta \rho'} = g_{\alpha \beta} (2k_1 - 2k_2 -l)_{\rho'} + g_{\beta \rho'} (k_2 - k_1 + 2l)_{\alpha}
+ g_{\rho' \alpha} (k_2 - k_1 -l)_{\beta}$, and we find that
only the terms proportional to $g_{\alpha \beta}$ and $g_{\rho' \alpha}$ contribute to the
LO hard kernel with $\gamma_{\alpha} = \gamma^-$.
Then we can factorize the amplitude $G^{(1)}_{2d,32}$ into the NLO twist-3 transversal rho DAs
$\phi^{(1),v}_{\rho, d} $ and $\phi^{(1),a}_{\rho, d}$
in Eq.~\ref{eq:2nlodrho}, convoluted with the LO hard amplitudes $G^{(0),v}_{a,32}(x_1; x_2) $ and
$G^{(0),a}_{a,32}(x_1; x_2) $, to which the gluon momenta flow or not flow in.

For Fig.~\ref{fig:fig2}(e) we have
\beq
G^{(1)}_{2e,32} = && \frac{i e g^4_s Tr[T^a T^c T^b] f_{abc}}{2 N_c}
    \frac{[\epsl_{1T} M_{\rho} \phi^v_{\rho} + i M_{\rho} \epsilon_{\mu' \nu \rho \sigma} \gamma_5 \gamma^{\mu'}
    \epsilon^{\nu}_{1T} n^{\rho} v^{\sigma} \phi^a_{\rho}]}{(p_1-k_2)^2 (k_1-k_2)^2 (k_1-l)^2 (k_1-k_2-l)^2 l^2} \non
 && \cdot \gamma^{\gamma} (\ksl_1 - \lsl) \gamma^{\alpha} [\gamma_5 \psl_2 \phi^A_{\pi}] \gamma_{\mu} (\psl_1 - \ksl_2)
    \gamma^{\beta} F_{\alpha \beta \gamma} \non
\sim &&0,
\label{eq:2nloe32}
\eeq
where $F_{\alpha \beta \gamma} = g_{\alpha \beta} (2k_1 - 2k_2 -l)_{\gamma} + g_{\beta \gamma}
(k_2 - k_1 - l)_{\alpha} + g_{\gamma \alpha} (k_2 - k_1 + 2l)_{\beta}$.
The possible contributions from the three terms in the tensor
$F_{\alpha \beta \gamma}$ is either suppressed by the kinetics or
excluded by the requirement that the Gamma matrix in the NLO amplitudes should hold
the LO content $\gamma_{\alpha} = \gamma^-$.
Then we can assume that the infrared contribution from the sub-diagram Fig.~\ref{fig:fig2}(e)
can be neglected safely.
The kinetic suppression is also happened for the amplitudes of Figs.~\ref{fig:fig2}(f,g),
theses two sub-diagrams also do not provide infrared correction to the LO hand kernel $G^{(0),v/a}_{a,32}$,
i.e.,
\beq
G^{(1)}_{2f,32} = && \frac{e g^4_s Tr[T^c T^a T^c T^a]}{2 N_c}
\frac{[\epsl_{1T} M_{\rho} \phi^v_{\rho} + i M_{\rho} \epsilon_{\mu' \nu \rho \sigma} \gamma_5 \gamma^{\mu'}
\epsilon^{\nu}_{1T} n^{\rho} v^{\sigma} \phi^a_{\rho}]}{(p_1-k_2)^2 (k_1-k_2)^2 (p_1-k_1+l)^2 (p_1-k_2+l)^2 l^2} \non
&& \cdot \gamma^{\alpha} [\gamma_5 \psl_2 \phi^A_{\pi}] \gamma_{\mu} (\psl_1 - \ksl_2) \gamma^{\rho'}
(\psl_1-\ksl_2+\lsl) \gamma_{\alpha} (\psl_1-\ksl_1+\lsl) \gamma_{\rho'} \non
\sim &&0,
\label{eq:2nlof32}
\eeq
\beq
G^{(1)}_{2g,32} = && \frac{- e g^4_s Tr[T^c T^a T^c T^a]}{2 N_c}
\frac{[\epsl_{1T} M_{\rho} \phi^v_{\rho} + i M_{\rho} \epsilon_{\mu' \nu \rho \sigma} \gamma_5 \gamma^{\mu'}
\epsilon^{\nu}_{1T} n^{\rho} v^{\sigma} \phi^a_{\rho}]}{(p_1-k_2)^2 (k_1-k_2-l)^2 (p_1-k_2-l)^2 (k_1-l)^2 l^2} \non
&& \cdot \gamma_{\rho'} (\ksl_1 - \lsl) \gamma^{\alpha} [\gamma_5 \psl_2 \phi^A_{\pi}] \gamma_{\mu} (\psl_1 - \ksl_2)
\gamma^{\rho'} (\psl_1-\ksl_2-\lsl) \gamma_{\alpha}\non
\sim &&0.
\label{eq:2nlog32}
\eeq

For sub-diagrams Figs.~\ref{fig:fig2}(h,i,j,k), however, the additional gluon
generates the collinear divergences as well as the soft divergences,
because both ends of the gluon are attached to the external quark lines.
As the partner with the soft divergences, the collinear divergences are also
evaluated by setting the radiative gluon momenta being parallel to
the initial rho meson momenta $p_1$.
The amplitudes for all these four sub-diagrams are given
in Eqs.~(\ref{eq:2nloh32},\ref{eq:2nloi32},\ref{eq:2nloj32},\ref{eq:2nlok32}).

For Figs.~\ref{fig:fig2}(h,i) we have
\beq
G^{(1)}_{2h,32} = && \frac{e g^4_s Tr[T^c T^a T^c T^a]}{2 N_c}
    \frac{[\epsl_{1T} M_{\rho} \phi^v_{\rho} + i M_{\rho} \epsilon_{\mu' \nu \rho \sigma} \gamma_5 \gamma^{\mu'}
    \epsilon^{\nu}_{1T} n^{\rho} v^{\sigma} \phi^a_{\rho}]}{(k_1-k_2)^2 (p_1-k_2+l)^2 (p_1-k_1+l)^2 (p_2-k_2+l)^2 l^2} \non
 && \cdot \gamma^{\alpha} [\gamma_5 \psl_2 \phi^A_{\pi}] \gamma^{\rho'} (\psl_2-\ksl_2+\lsl) \gamma_{\mu} (\psl_1-\ksl_2+\lsl)
    \gamma^{\alpha} (\psl_1-\ksl_1+\lsl) \gamma_{\rho'} \non
\sim && (-\frac{1}{8}) \phi^{(1),v}_{\rho, d} \otimes G^{(0),v}_{a,32}(x_1, x_2)
      + (-\frac{1}{8}) \phi^{(1),a}_{\rho, d} \otimes G^{(0),a}_{a,32}(x_1, x_2),
\label{eq:2nloh32}
\eeq
\beq
G^{(1)}_{2i,32} = && \frac{-e g^4_s Tr[T^c T^a T^c T^a]}{2 N_c}
    \frac{[\epsl_{1T} M_{\rho} \phi^v_{\rho} + i M_{\rho} \epsilon_{\mu' \nu \rho \sigma} \gamma_5 \gamma^{\mu'}
    \epsilon^{\nu}_{1T} n^{\rho} v^{\sigma} \phi^a_{\rho}]}{(k_1-k_2-l)^2 (p_1-k_2)^2 (p_1-k_1+l)^2 (k_2+l)^2 l^2} \non
 && \cdot \gamma^{\alpha} (\ksl_2+\lsl) \gamma^{\rho'} [\gamma_5 \psl_2 \phi^A_{\pi}] \gamma_{\mu} (\psl_1-\ksl_2)
    \gamma_{\alpha} (\psl_1-\ksl_1+\lsl) \gamma_{\rho'} \non
\sim && (\frac{1}{8}) \phi^{(1),v}_{\rho, d} \otimes G^{(0),v}_{a,32}(\xi_1; x_2)
      + (\frac{1}{8}) \phi^{(1),a}_{\rho, d} \otimes G^{(0),a}_{a,32}(\xi_1; x_2).
\label{eq:2nloi32}
\eeq

For Figs.~\ref{fig:fig2}(j,k), we find that $G^{(1)}_{2j,32}$ and $G^{(1)}_{2k,32}$
don't provide the NLO correction to the LO amplitude $G^{(0)}_{a,32}$,
because of the confine of the Gamma matrixes to extract the LO amplitude
$G^{(0)}_{a,32}$,
then the infrared contribution of these two amplitudes can also be neglected safely.
\beq
G^{(1)}_{2j,32} = && \frac{e g^4_s Tr[T^c T^a T^c T^a]}{2 N_c}
\frac{[\epsl_{1T} M_{\rho} \phi^v_{\rho} + i M_{\rho} \epsilon_{\mu' \nu \rho \sigma} \gamma_5 \gamma^{\mu'}
\epsilon^{\nu}_{1T} n^{\rho} v^{\sigma} \phi^a_{\rho}]}{(k_1-k_2)^2 (p_1-k_2)^2 (k_1-l)^2 (k_2-l)^2 l^2} \non
&& \cdot \gamma_{\rho'} (\ksl_1-\lsl) \gamma_{\bot}^{\alpha} (\ksl_2-\lsl) \gamma^{\rho'}
[\gamma_5 \psl_2 \phi^A_{\pi}] \gamma_{\mu} (\psl_1-\ksl_2) \gamma_{\bot \alpha}\non
\sim && 0,
\label{eq:2nloj32}
\eeq
\beq
G^{(1)}_{2k,32} = && \frac{- e g^4_s Tr[T^c T^a T^c T^a]}{2 N_c}
\frac{[\epsl_{1T} M_{\rho} \phi^v_{\rho} + i M_{\rho} \epsilon_{\mu' \nu \rho \sigma} \gamma_5 \gamma^{\mu'}
\epsilon^{\nu}_{1T} n^{\rho} v^{\sigma} \phi^a_{\rho}]}{(k_1-k_2-l)^2 (p_1-k_2-l)^2 (k_1-l)^2 (p_2-k_2-l)^2 l^2} \non
&& \cdot \gamma_{\rho'} (\ksl_1-\lsl) \gamma_{\bot}^{\alpha} [\gamma_5 \psl_2 \phi^A_{\pi}] \gamma^{\rho'} (\psl_2-\ksl_2-\lsl)
 \gamma_{\mu} (\psl_1-\ksl_2-\lsl) \gamma_{\bot \alpha}\non
\sim && 0,
\label{eq:2nlok32}
\eeq

For the irreducible infrared amplitudes as shown in
Eqs.~(\ref{eq:2nlod32},\ref{eq:2nloe32}-\ref{eq:2nlok32}), we have the following observations:
\begin{enumerate}
\item[(i)]
We sum up the amplitudes for the irreducible sub-diagrams Figs.~\ref{fig:fig2}(d,f,h,i) together,
in which the additional gluon is radiated from the initial up-line quark.
\beq
G^{(1)}_{2up,32} && (x_1; x_2) = G^{(1)}_{2d,32}(x_1; x_2) + G^{(1)}_{2f,32}(x_1; x_2)
                                 + G^{(1)}_{2h,32}(x_1; x_2) + G^{(1)}_{2i,32}(x_1; x_2) \non
  =&& \phi^{(1),v}_{\rho,d} \otimes \left (\frac{7}{16}\right )
  \left [G^{(0),v}_{a,32}(x_1; x_2) - G^{(0),v}_{a,32}(\xi_1; x_2) \right ] \non
  +&& \phi^{(1),a}_{\rho,d} \otimes \left (\frac{7}{16}\right )
\left [G^{(0),a}_{a,32}(x_1; x_2) - G^{(0),a}_{a,32}(\xi_1; x_2) \right ] .
\label{eq:2nloup32}
\eeq
The summation of the amplitudes for the sub-diagrams Figs.~\ref{fig:fig2}(e,g,j,k),
in which the additional gluon is radiated from the initial down-line quark,
would give the zero infrared contribution.
The infrared divergences only come from the gluon radiated from the up-line quark
of rho meson as shown in Fig.~\ref{fig:fig2},
while the infrared contributions from the down-line quark are excluded either
by the dynamics or the kinetics.

\item[(ii)]
By comparing the amplitudes $G^{(1)}_{2h,32}$ with $G^{(1)}_{2i,32}$,
We find that the soft divergences from the irreducible sub-diagrams
Fig.~\ref{fig:fig2}(h) and Fig.~\ref{fig:fig2}(i) will  be canceled completely
by the simple replacement $\xi_1 \to x_1$.
Combining with the cancellation of the soft divergences in the sub-diagrams
Figs.~\ref{fig:fig2}(a,b,c),  there is no soft divergence in the quark level
for the Fig.~\ref{fig:fig2}.

\item[(iii)]
The NLO corrections to the LO sub-diagram Fig.~\ref{fig:fig1}(a) with the collinear
gluon emitted from the initial state do have the collinear divergences, but they
can be absorbed into the NLO rho meson DAs $\phi^{(1),v}_{\rho,d}$ and $\phi^{(1),a}_{\rho,d}$.
From Eqs.~(\ref{eq:2nlod32},\ref{eq:2nlof32},\ref{eq:2nloh32},\ref{eq:2nloi32}),
one can write out the Feynman rules for the perturbative calculation of the NLO twist-3
transversal $\rho$ meson wave functions
  $\phi^{(1),v}_{\rho,d}$ and $\phi^{(1),a}_{\rho,d}$ as a nonlocal hadronic matrix
element with the structure
  $\gamma_{\bot}/2$ and $(\gamma_5 \gamma_{\bot})/2$ sandwiched respectively:
{\small
\beq
\phi^{(1),v}_{\rho,d}&=&\frac{1}{2N_c P^+_1} \int \frac{dy^-}{2\pi} e^{-i x p^+_1 y^-}
    <0| \overline{q}(y^-) \frac{\gamma_{\bot}}{2} (-i g_s)  \int^{y^-}_{0} dz v
\cdot A(zv) q(0)| \rho(p_1)>,
  \label{eq:nlorhovd}\\
\phi^{(1),a}_{\rho,d}&=&\frac{1}{2N_c P^+_1} \int \frac{dy^-}{2\pi} e^{-i x p^+_1 y^-}
    <0| \overline{q}(y^-) \frac{\gamma_5 \gamma_{\bot}}{2} (-i g_s) \int^{y^-}_{0} dz v \cdot A(zv) q(0)| \rho(p_1)>.\ \
  \label{eq:nlorhoad}
\eeq }
The integral variable $z$ runs from $0$ to $\infty$ for the upper eikonal line as showed in Fig.~\ref{fig:fig3}(a),
and runs from $\infty$ back to $y^-$ for the lower eikonal line  as showed in Fig.~\ref{fig:fig3}(b).
The choice of the light-cone coordinate $y^- \neq 0$ represents the fact
that the collinear divergences from the sub-diagrams of Fig.~\ref{fig:fig2}  don't cancel
exactly.

  \begin{figure}[tb]
  \centering
  \vspace{-1cm}
  \begin{center}
  \leftline{\epsfxsize=12cm\epsffile{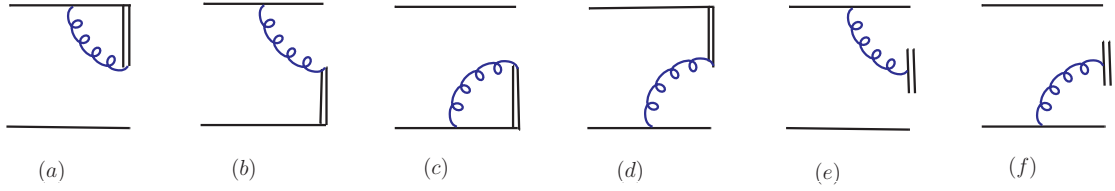}}
  \end{center}
  \vspace{-14cm}
  \caption{${\calo}(\alpha_s)$ effective diagrams for the initial transversal rho meson wave function,
  which collect all the collinear divergences from the initial rho meson in the irreducible NLO quark diagrams.
  The vertical double line denotes the Wilson line along the light cone, whose Feynman rule is
  $v_\rho/(v \cdot l)$ as described in Eq.~(\ref{eq:feyn-wilsonline1},\ref{eq:feyn-wilsonline2}).}
  \label{fig:fig3}
  \end{figure}
\item[(iv)]
The factor $v_\rho/(v\cdot l$) in Eq.~(\ref{eq:2nlodrho}) is the Feynman rule associated with the Wilson line,
which is required to remain gauge invariance of the nonlocal matrix element in the NLO rho wave functions
and has been included in Eqs.~(\ref{eq:nlorhovd},\ref{eq:nlorhoad}) of the NLO wave functions.
We can retrieve this factor by Fourier transformation of the gauge field from $A(zv)$ to $A(l)$ in these NLO wave functions:
\beq
\int^{\infty}_{0} dz v\cdot A(zv)
&& \to \int dl ~e^{iz(v \cdot l + i\epsilon)}~\int^{\infty}_{0} dz v\cdot \widetilde{A}(l)
= i \int dl ~\frac{v_\rho}{v\cdot l} \widetilde{A}^\rho(l),
\label{eq:feyn-wilsonline1} \\
\int^{y^-}_{0} dz v\cdot A(zv)
&& \to \int dl ~e^{iz(v \cdot l + i\epsilon)}~\int^{y^-}_{0} dz v\cdot \widetilde{A}(l)
= -i \int dl ~\frac{v_\rho}{v\cdot l} ~e^{il^+y^-}~\widetilde{A}^\rho(l).
\label{eq:feyn-wilsonline2}
\eeq
The Fourier factor $e^{il^+y^-}$ in Eq.~(\ref{eq:feyn-wilsonline2}) will lead to the function
$\delta(\zeta_1-x_1+i^+/p^+_1)$, which means that the gluon momentum $l$ has flowed into
the LO hard kernel as described in Eqs.~(\ref{eq:loa32iv},\ref{eq:loa32ia}).

\item[(v)]
The NLO irreducible amplitudes for Fig.~\ref{fig:fig2} in the collinear region
can be written as the convolutions of the NLO DAs and the LO hard amplitudes.
The collinear factorization is valid for the NLO corrections for the
Fig.~\ref{fig:fig1}(a)  with the additional gluon emitted from the initial rho meson.

\item[(vi)]
The sub-diagrams Figs.~\ref{fig:fig3}(a,b,e) are the effective-diagrams for the additional
gluon radiated from the left-up quark line, the sub-diagrams Figs.~\ref{fig:fig3}(c,d,f)
represent the effective-diagrams for the additional gluon radiated from the left-down
anti-quark line.
We can also sort these six effective-diagrams in Fig.~3 into three sets by the
flowing of the gluon  momenta:
(a) the first set contains the effective diagram 3(a) and 3(c) with no gluon momenta flow
into the LO hard amplitudes;
(b) the second set is made of the effective diagram 3(b) and 3(d) with the gluon
momenta flow into the  LO hard amplitudes;
and (c) the third set includes  the effective diagram 3(e) and 3(f) with the
gluon momenta flow partly into the LO hard amplitudes.

\end{enumerate}

Now we consider the infrared divergences from $\calo(\alpha_s)$ radiative corrections to
Fig.~\ref{fig:fig1}(a) with the additional collinear gluon emitted from the final $\pi$ meson
as shown in Fig.~\ref{fig:fig4},
where the gluon momenta may be collinear with the pion meson momenta $p_2$.

Since the sub-diagrams Figs.~\ref{fig:fig4}(a,b,c) are reducible diagrams,
we can factorize them directly  by inserting the Firez identity into proper places
as being done for Figs.~\ref{fig:fig2}(a,b,c) previously.
The symmetry factor $1/2$ are also exist in $G^{(1)}_{4a,32}$ and $G^{(1)}_{4c,32}$.
And the soft divergences in these reducible amplitudes $G^{(1)}_{4a,32}, G^{(1)}_{4b,32},
G^{(1)}_{4c,32}$ as given in Eqs.~(\ref{eq:4nloa32}.\ref{eq:4nlob32},\ref{eq:4nloc32})
will also be cancelled each other exactly.

\begin{figure}[tb]
\vspace{-1cm}
\begin{center}
\leftline{\epsfxsize=10cm\epsffile{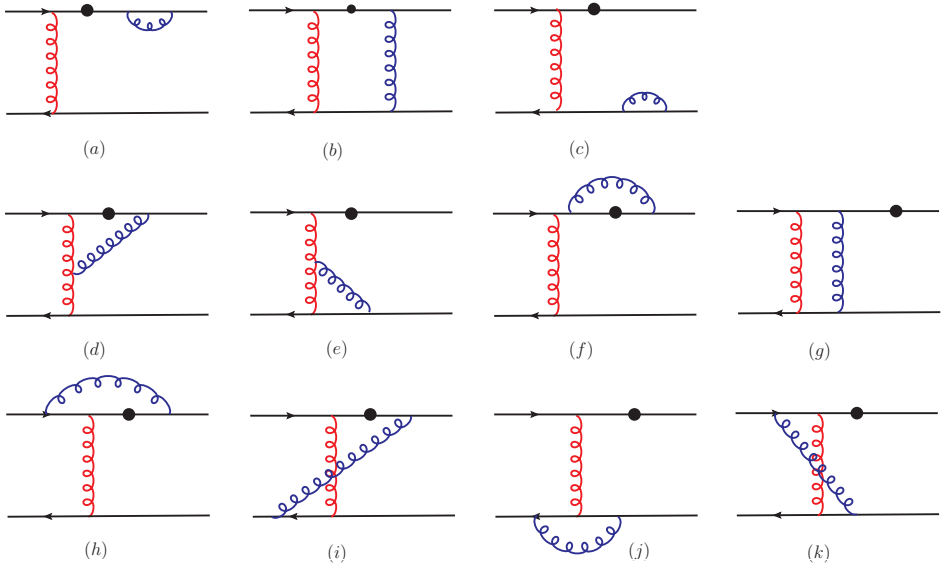}}
\end{center}
\vspace{-6cm}
\caption{$\calo(\alpha_s)$ corrections to Fig.~\ref{fig:fig1}(a) with an additional gluon
(blue curves) emitted from the final $\pi$ meson.}
\label{fig:fig4}
\end{figure}
\beq
G^{(1)}_{4a,32} =&& \frac{1}{2} \frac{e g^4_s C^2_F}{2}
    \frac{[\epsl_{1T} M_{\rho} \phi^v_{\rho} + i M_{\rho} \epsilon_{\mu' \nu \rho \sigma} \gamma_5 \gamma^{\mu'}
    \epsilon^{\nu}_{1T} n^{\rho} v^{\sigma} \phi^a_{\rho}]\gamma^{\alpha} [\gamma_5 \psl_2 \phi^A_{\pi}] \gamma^{\rho'}}
    {(p_1-k_2)^2 (k_1-k_2)^2 (p_2-k_2)^2 (p_2-k_2+l)^2 l^2} \non
 && \cdot (\psl_2-\ksl_2+\lsl) \gamma_{\rho'}
    (\psl_2-\ksl_2) \gamma_{\mu} (\psl_1 - \ksl_2) \gamma_{\alpha} \non
=&& \frac{1}{2} G^{(0)}_{a,32}(x_1; x_2) \otimes \phi^{(1),A}_{\pi, a},
\label{eq:4nloa32}
\eeq
\beq
G^{(1)}_{4b,32} =&& \frac{- e g^4_s C^2_F}{2}
    \frac{[\epsl_{1T} M_{\rho} \phi^v_{\rho} + i M_{\rho} \epsilon_{\mu' \nu \rho \sigma} \gamma_5 \gamma^{\mu'}
    \epsilon^{\nu}_{1T} n^{\rho} v^{\sigma} \phi^a_{\rho}]}{(p_1-k_2+l)^2 (k_1-k_2+l)^2 (p_2-k_2+l)^2 (k_2-l)^2 l^2} \non
 && \cdot \gamma^{\rho'} [\gamma_5 \psl_2 \phi^A_{\pi}] \gamma_{\rho'} (\psl_2-\ksl_2+\lsl) \gamma_{\mu} (\psl_1-\ksl_2+\lsl)
    \gamma_{\alpha} \non
=&& G^{(0)}_{a,32}(x_1; \xi_2) \otimes \phi^{(1),A}_{\pi, b},
\label{eq:4nlob32}
\eeq
\beq
G^{(1)}_{4c,32} =&& \frac{1}{2} \frac{e g^4_s C^2_F}{2}
    \frac{[\epsl_{1T} M_{\rho} \phi^v_{\rho} + i M_{\rho} \epsilon_{\mu' \nu \rho \sigma} \gamma_5 \gamma^{\mu'}
    \epsilon^{\nu}_{1T} n^{\rho} v^{\sigma} \phi^a_{\rho}]\gamma^{\alpha} \ksl_2 \gamma^{\rho'} (\ksl_2-\lsl)}
    {(p_1-k_2)^2 (k_1-k_2)^2 (k_2-l)^2 (k_2)^2 l^2} \non
 && \cdot \gamma_{\rho'}[\gamma_5 \psl_2 \phi^A_{\pi}] \gamma_{\mu} (\psl_1-\ksl_2) \gamma_{\alpha} \non
=&& \frac{1}{2} G^{(0)}_{a,32}(x_1, x_2) \otimes \phi^{(1),A}_{\pi, c},
\label{eq:4nloc32}
\eeq
where $\phi^{(1),A}_{\pi, i}$ with $i=(a,b,c)$ are the NLO DAs, which absorbed all the
infrared singularities from these reducible sub-diagrams Figs.~\ref{fig:fig2}(a,b,c)
and can be written in the following forms:
\beq
\phi^{(1),A}_{\pi, a} &&= \frac{- i g^2_s C_F}{4} \frac{[\gamma_5 \gamma^+] \gamma^{\rho'} (\psl_2-\ksl_2+\lsl)
    \gamma_{\rho'} (\psl_2-\ksl_2) [\gamma^- \gamma_5]}{(p_2-k_2)^2 (p_2-k_2+l)^2 l^2}; \non
\phi^{(1),A}_{\pi, b} &&= \frac{i g^2_s C_F}{4} \frac{(\ksl_2-\lsl) \gamma^{\rho'} [\gamma_5 \gamma^+] \gamma_{\rho'}
    (\psl_2-\ksl_2+\lsl) [\gamma^- \gamma_5]}{(p_2-k_2+l)^2 (k_2-l)^2 l^2}; \non
\phi^{(1),A}_{\pi, c} &&= \frac{- i g^2_s C_F}{4} \frac{[\gamma^- \gamma_5] \ksl_2 \gamma^{\rho'} (\ksl_2-\lsl)
    \gamma_{\rho'} [\gamma_5 \gamma^+]}{(k_2-l)^2 (k_2)^2 l^2}.
\label{eq:4nloabcpionA}
\eeq

The infrared singularity analysis for Fig.~\ref{fig:fig2} are also valid for Fig.~\ref{fig:fig4}.
The sub-diagrams in the second row of Fig.~\ref{fig:fig4} also contain the collinear singularity only,
while the third row sub-diagrams may contain both collinear and soft divergences.
Before discussing the infrared behaviour of these irreducible sub-diagrams in
Figs.~\ref{fig:fig4}(d-k), we here firstly define those LO hard amplitudes which either
appeared in Eq.~\ref{eq:4nlob32} or will appeare  in the NLO irreducible amplitudes,
\beq
G^{(0)}_{a,32}(x_1;\xi_2)= &&\frac{i e g^2_s C_F}{2} \frac{[\epsl_{1T} M_{\rho} \phi^{v}_{\rho}
    + M_{\rho} i \epsilon_{\mu'\nu\rho\sigma} \gamma_5 \gamma^{\mu'} \epsilon^{\nu}_{1T} n^{\rho} v^{\sigma}]}
    {(p_1-k_2+l)^2 (k_1-k_2+l)^2} \non
 && \cdot \gamma^{\alpha} [\gamma_5 \psl_2 \phi^A_{\pi}] \gamma_{\mu} (\psl_1 - \ksl_2 +\lsl) \gamma_{\alpha},
\label{eq:loa32fA1} \\
G^{(0)}_{a,32}(x_1; \xi_2, x_2)= &&\frac{i e g^2_s}{2} \frac{[\epsl_{1T} M_{\rho} \phi^v_{\rho}
    + i M_{\rho} \epsilon_{\mu' \nu \rho \sigma}\gamma_5 \gamma^{\mu'} \epsilon^{\nu}_{1T} n^{\rho} v^{\sigma} \phi^a_{\rho}]}{(p_1-k_2+l)^2 (k_1-k_2)^2} \non
 && \cdot \gamma^{\alpha} [\gamma_5 \psl_2 \phi^A_{\pi}] \gamma_{\mu} (\psl_1-\ksl_2) \gamma_{\alpha},
\label{eq:loa32fA2} \\
G^{'(0)}_{a,32}(x_1; \xi_2, x_2)= &&\frac{i e g^2_s}{2} \frac{[\epsl_{1T} M_{\rho} \phi^v_{\rho}
    + i M_{\rho} \epsilon_{\mu' \nu \rho \sigma}\gamma_5 \gamma^{\mu'} \epsilon^{\nu}_{1T} n^{\rho} v^{\sigma} \phi^a_{\rho}]}{(p_1-k_2)^2 (k_1-k_2+l)^2} \non
 && \cdot \gamma^{\alpha} [\gamma_5 \psl_2 \phi^A_{\pi}] \gamma_{\mu} (\psl_1-\ksl_2) \gamma_{\alpha}.
\label{eq:loa32fA3}
\eeq

In the collinear region $l \parallel p_2$, we can find the equal relation
$G^{(0)}_{a,32}(x_1; \xi_2, x_2)=G^{'(0)}_{a,32}(x_1; \xi_2, x_2)$
for the newly defined LO hard amplitudes as shown in Eqs.~(\ref{eq:loa32fA2},\ref{eq:loa32fA3}).

The transition amplitude for Fig.~\ref{fig:fig4}(d) can be written as the form of
\beq
G^{(1)}_{4d,32} =&& \frac{- i e g^4_s Tr[T^c T^b T^a] f_{abc}}{2 N_c}
    \frac{[\epsl_{1T} M_{\rho} \phi^v_{\rho} + i M_{\rho} \epsilon_{\mu' \nu \rho \sigma} \gamma_5 \gamma^{\mu'}
    \epsilon^{\nu}_{1T} n^{\rho} v^{\sigma} \phi^a_{\rho}]}{(p_1-k_2+l)^2 (k_1-k_2)^2 (k_1-k_2+l)^2 (p_2-k_2+l)^2 l^2} \non
 && \cdot \gamma^{\alpha} [\gamma_5 \psl_2 \phi^A_{\pi}] \gamma^{\beta} (\psl_2-\ksl_2+\lsl) \gamma_{\mu} (\psl_1-\ksl_2+\lsl)
    \gamma^{\gamma} F_{\alpha \beta \gamma}\non
=&& \left [G^{(0)}_{a,32}(x_1; \xi_2, x_2) - G^{(0)}_{a,32}(x_1; \xi_2)\right]
\otimes \frac{9}{16} \phi^{(1),A}_{\pi, d},
\label{eq:4nlod32}
\eeq
with the tensor $F_{\alpha \beta \gamma} = g_{\alpha \beta} (k_1-k_2-l)_{\gamma} + g_{\beta \gamma} (k_1-k_2+2l)_{\alpha}
+g_{\gamma \alpha} (2k_2-2k_1-l)_{\beta}$,
in which only terms proportional to $g_{\beta \gamma}$ and $g_{\gamma \alpha}$
contribute to the LO hard kernel $G^{(0)}_{a,32}$.
The NLO twist-2 pion DA $\phi^{(1),A}_{\pi, d}$ is defined in the following form
\beq
\phi^{(1),A}_{\pi, d} = \frac{- i g^2_s C_F}{4}
    \frac{[\gamma_5 \gamma^+] \gamma^{\rho} (\psl_2-\ksl_2+\lsl) [\gamma^- \gamma_5] n_{\rho'}}
    {(p_2-k_2+l)^2 l^2 (n\cdot l)}.
\label{eq:4nlodpionA}
\eeq
Here the eikonal approximation has been employed to obtain
the convolution forms for these irreducible amplitudes.

For Fig.~\ref{fig:fig4}(e), similarly, we have
\beq
G^{(1)}_{4e,32} =&& \frac{i e g^4_s Tr[T^c T^b T^a] f_{abc}}{2 N_c}
    \frac{[\epsl_{1T} M_{\rho} \phi^v_{\rho} + i M_{\rho} \epsilon_{\mu' \nu \rho \sigma} \gamma_5 \gamma^{\mu'}
    \epsilon^{\nu}_{1T} n^{\rho} v^{\sigma} \phi^a_{\rho}]}{(p_1-k_2)^2 (k_1-k_2)^2 (k_1-k_2+l)^2 (k_2-l)^2 l^2} \non
 && \cdot \gamma^{\alpha} (\ksl_2-\lsl) \gamma^{\beta} [\gamma_5 \psl_2 \phi^A_{\pi}] \gamma_{\mu} (\psl_1-\ksl_2)
    \gamma^{\gamma} F_{\alpha \beta \gamma}\non
=&& \left [G^{(0)}_{a,32}(x_1; x_2) - G^{'(0)}_{a,32}(x_1; \xi_2, x_2) \right]
\otimes \frac{9}{16} \phi^{(1),A}_{\pi, e},
\label{eq:4nloe32}
\eeq
where $F_{\alpha \beta \gamma} = g_{\alpha \beta} (k_1-k_2+2l)_{\gamma} + g_{\beta \gamma} (k_1-k_2-l)_{\alpha}
+g_{\gamma \alpha} (2k_2-2k_1-l)_{\beta}$, and only the terms proportional to
$g_{\beta \gamma}$ and $g_{\gamma \alpha}$ contribute to the LO hard kernel $G^{(0)}_{a,32}$.
The NLO twist-2 pion DA $\phi^{(1),A}_{\pi, e}$ is defined in the form of
\beq
\phi^{(1),A}_{\pi, e} = \frac{i g^2_s C_F}{4}
    \frac{[\gamma_5 \gamma^+] \gamma^{\rho} (\ksl_2-\lsl) [\gamma^- \gamma_5] n_{\rho'}}
    {(k_2-l)^2 l^2 (n\cdot l)},
\label{eq:4nloepionA}
\eeq
where the additional gluon is emitted from the right-down anti-parton line.
Then the amplitudes for the remaining irreducible sub-diagrams in Fig.~\ref{fig:fig4}
can be written with the definitions in
Eqs.~(\ref{eq:loa32fA1},\ref{eq:loa32fA2},\ref{eq:loa32fA3},\ref{eq:4nlodpionA},\ref{eq:4nloepionA}):
\beq
G^{(1)}_{4f,32} =&& \frac{e g^4_s C^2_F N_c}{2 N_c}
    \frac{[\epsl_{1T} M_{\rho} \phi^v_{\rho} + i M_{\rho} \epsilon_{\mu' \nu \rho \sigma} \gamma_5 \gamma^{\mu'}
    \epsilon^{\nu}_{1T} n^{\rho} v^{\sigma} \phi^a_{\rho}]}{(p_1-k_2)^2 (k_1-k_2)^2 (p_2-k_2+l)^2 l^2 (p_1-k_2+l)^2} \non
 && \cdot \gamma^{\alpha} [\gamma_5 \psl_2 \phi^A_{\pi}] \gamma^{\rho'} (\psl_2-\ksl_2+\lsl) \gamma_{\mu} (\psl_1-\ksl_2+\lsl)
    \gamma_{\rho'} (\psl_1-\ksl_2) \gamma_{\alpha}\non
=&& \left [G^{(0)}_{a,32}(x_1; x_2) - G^{(0)}_{a,32}(x_1; \xi_2, x_2) \right]
\otimes \phi^{(1),A}_{\pi, d},
\label{eq:4nlof32}
\eeq
\beq
G^{(1)}_{4g,32} = &&\frac{- e g^4_s C^2_F N_c}{2 N_c}
    \frac{[\epsl_{1T} M_{\rho} \phi^v_{\rho} + i M_{\rho} \epsilon_{\mu' \nu \rho \sigma} \gamma_5 \gamma^{\mu'}
    \epsilon^{\nu}_{1T} n^{\rho} v^{\sigma} \phi^a_{\rho}]}{(p_1-k_2)^2 (k_1-k_2+l)^2 (k_2-l)^2 l^2 (p_1-k_2+l)^2} \non
 && \cdot \gamma^{\alpha} (\ksl_2-\lsl) \gamma^{\rho'} [\gamma_5 \psl_2 \phi^A_{\pi}] \gamma_{\mu} (\psl_1-\ksl_2)
    \gamma_{\rho'} (\psl_1-\ksl_2+\lsl) \gamma_{\alpha} \non
=&& \left [G^{'(0)}_{a,32}(x_1; \xi_2, x_2) - G^{(0)}_{a,32}(x_1; \xi_2) \right]
\otimes \phi^{(1),A}_{\pi, e},
\label{eq:4nlog32}
\eeq
\beq
G^{(1)}_{4h,32} =&& \frac{e g^4_s Tr[T^c T^a T^c T^a]}{2 N_c}
    \frac{[\epsl_{1T} M_{\rho} \phi^v_{\rho} + i M_{\rho} \epsilon_{\mu' \nu \rho \sigma} \gamma_5 \gamma^{\mu'}
    \epsilon^{\nu}_{1T} n^{\rho} v^{\sigma} \phi^a_{\rho}]}{(p_1-k_2+l)^2 (k_1-k_2)^2 (p_2-k_2+l)^2 l^2 (p_1-k_1+l)^2} \non
 && \cdot \gamma^{\alpha} [\gamma_5 \psl_2 \phi^A_{\pi}] \gamma^{\rho'} (\psl_2-\ksl_2+\lsl) \gamma_{\mu} (\psl_1-\ksl_2+\lsl)
    \gamma_{\alpha} (\psl_1-\ksl_1+\lsl) \gamma_{\rho'} \non
=&& G^{(0)}_{a,32}(x_1; \xi_2, x_2) \otimes (-\frac{1}{8})\phi^{(1),A}_{\pi, d},
\label{eq:4nloh32}
\eeq
\beq
G^{(1)}_{4i,32} =&& \frac{- e g^4_s Tr[T^c T^a T^c T^a]}{2 N_c}
    \frac{[\epsl_{1T} M_{\rho} \phi^v_{\rho} + i M_{\rho} \epsilon_{\mu' \nu \rho \sigma} \gamma_5 \gamma^{\mu'}
    \epsilon^{\nu}_{1T} n^{\rho} v^{\sigma} \phi^a_{\rho}]}{(p_1-k_2+l)^2 (k_1-k_2+l)^2 (p_2-k_2+l)^2 l^2 (k_1+l)^2} \non
 && \cdot \gamma_{\rho'} (\ksl_1+\lsl) \gamma^{\alpha} [\gamma_5 \psl_2 \phi^A_{\pi}] \gamma^{\rho'} (\psl_2-\ksl_2+\lsl)
    \gamma_{\mu} (\psl_1-\ksl_2+\lsl) \gamma_{\alpha} \non
=&& 0,
\label{eq:4nloi32}
\eeq
\beq
G^{(1)}_{4j,32} =&& \frac{e g^4_s Tr[T^c T^a T^c T^a]}{2 N_c}
    \frac{[\epsl_{1T} M_{\rho} \phi^v_{\rho} + i M_{\rho} \epsilon_{\mu' \nu \rho \sigma} \gamma_5 \gamma^{\mu'}
    \epsilon^{\nu}_{1T} n^{\rho} v^{\sigma} \phi^a_{\rho}]\gamma_{\rho'} (\ksl_1-\lsl) \gamma^{\alpha}}
    {(p_1-k_2)^2 (k_1-k_2)^2 (k_2-l)^2 l^2 (k_1-l)^2} \non
 && \cdot (\ksl_2-\lsl) \gamma^{\rho'} [\gamma_5 \psl_2 \phi^A_{\pi}]
    \gamma_{\mu} (\psl_1-\ksl_2) \gamma_{\alpha} \non
=&& 0,
\label{eq:4nloj32}
\eeq
\beq
G^{(1)}_{4k,32} =&& \frac{- e g^4_s Tr[T^c T^a T^c T^a]}{2 N_c}
    \frac{[\epsl_{1T} M_{\rho} \phi^v_{\rho} + i M_{\rho} \epsilon_{\mu' \nu \rho \sigma} \gamma_5 \gamma^{\mu'}
    \epsilon^{\nu}_{1T} n^{\rho} v^{\sigma} \phi^a_{\rho}]}{(p_1-k_2)^2 (k_1-k_2+l)^2 (k_2-l)^2 l^2 (p_1-k_1-l)^2} \non
 && \cdot \gamma^{\alpha} (\ksl_2-\lsl) \gamma^{\rho'} [\gamma_5 \psl_2 \phi^A_{\pi}] \gamma_{\mu} (\psl_1-\ksl_2)
    \gamma_{\alpha} (\psl_1-\ksl_1-\lsl) \gamma_{\rho'} \non
=&& G^{'(0)}_{a,32}(x_1; \xi_2, x_2) \otimes (\frac{1}{8}) \phi^{(1),A}_{\pi, e}.
\label{eq:4nlok32}
\eeq
The infrared contributions from the NLO amplitudes $G^{(1)}_{4j,32}$ and $G^{(1)}_{4j,32}$
are zero, since the Gamma matrixes in these two amplitudes are
$\gamma^{\alpha} = \gamma^{\alpha}_{\bot}$ instead of the $\gamma^{\alpha} = \gamma^-$ for the LO
amplitudes.

In order to investigate the NLO collinear factorization of the Fig.~\ref{fig:fig4} and to
extracte the NLO twist-2 pion meson DA, we make the summation over all the irreducible
amplitudes in Fig.~\ref{fig:fig4} into two sets:
the first set includes the sub-diagrams with the gluon radiated from the right-up quark
line of the final pion meson, while the second set containes the sub-diagrams with the gluon
radiated from the right-down quark line.

We firstly sum up the infrared amplitudes for the irreducible sub-diagrams in
Figs.~\ref{fig:fig4}(d,f,h,i) with the gluon radiated from the right-up quark line:
\beq
G^{(1)}_{\rm 4up,32} && (x_1; x_2) = G^{(1)}_{4d,32}(x_1; x_2) + G^{(1)}_{4f,32}(x_1; x_2)
                                 + G^{(1)}_{4h,32}(x_1; x_2) + G^{(1)}_{4i,32}(x_1; x_2) \non
  =&& \left [G^{(0)}_{a,32}(x_1; x_2) - \frac{9}{16}G^{(0)}_{a,32}(x_1; \xi_2)
- \frac{9}{16}G^{(0)}_{a,32}(x_1; x_2, \xi_2) \right ]
      \otimes \phi^{(1),A}_{\pi,d};
\label{eq:4nloup32}
\eeq

For the second set of  the irreducible sub-diagrams in Figs.~\ref{fig:fig4}(e,g,j,k)(
where  the gluon radiated from the right-down anti-quark line), similarly,
we make the summation and then find the infrared amplitude:
\beq
G^{(1)}_{\rm 4down,32} && (x_1; x_2) = G^{(1)}_{4e,32}(x_1; x_2) + G^{(1)}_{4h,32}(x_1; x_2)
                                   + G^{(1)}_{4j,32}(x_1; x_2) + G^{(1)}_{4k,32}(x_1; x_2) \non
  =&& \left [- G^{(0)}_{a,32}(x_1; \xi_2) + \frac{9}{16} G^{(0)}_{a,32}(x_1; x_2)
+ \frac{9}{16} G^{(0)}_{a,32}(x_1; x_2, \xi_2)\right ]
      \otimes \phi^{(1),A}_{\pi,e}.
  \label{eq:4nlodown32}
  \eeq

Because the IR singularities in Eqs.~(\ref{eq:4nloi32},\ref{eq:4nloj32}) are suppressed,
then the soft divergences in Eq.~(\ref{eq:4nloh32}) and Eq.~(\ref{eq:4nlok32}) from the
collinear region can't be cancelled by their counterparts described in Eq.~(\ref{eq:4nloi32})
and Eq.~(\ref{eq:4nloj32}) respectively.
But these remained soft divergences in Eqs.~(\ref{eq:4nloh32},\ref{eq:4nlok32}) could
be canceled each other exactly,
because the NLO DA $\phi^{(1),A}_{\pi,d}$ in Eq.~(\ref{eq:4nlodpionA})
is equivalent to the DA $\phi^{(1),A}_{\pi,e}$ in  Eq.~(\ref{eq:4nloepionA}).
At the quark level, finally, no soft divergences are left after summation of the NLO contributions
from all the sub-diagrams as shown in Fig.~\ref{fig:fig4}.

\begin{figure}[tb]
\vspace{-1cm}
\begin{center}
\leftline{\epsfxsize=12cm\epsffile{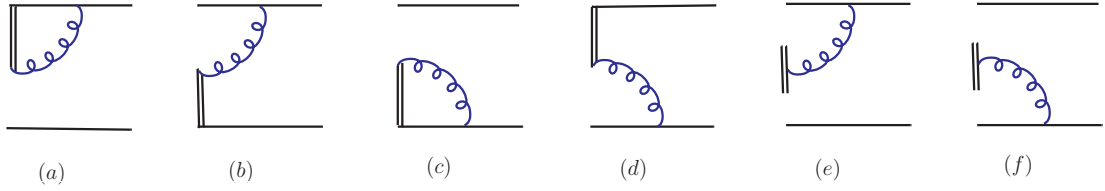}}
\end{center}
\vspace{-14cm}
\caption{${\calo}(\alpha_s)$ effective diagrams for the final pion meson wave function,
with vertical double line denoting the Wilson line along the light cone, whose Feynman rule is
$n_{\rho'}/(n \cdot l)$.}
\label{fig:fig5}
\end{figure}

All the remaining collinear divergences can be absorbed into the NLO twist-2 pion meson DA $\phi^{(1),A}_{\pi}$.
From the expressions as given in
Eqs.~(\ref{eq:4nlod32},\ref{eq:4nloe32},\ref{eq:4nlof32},\ref{eq:4nlog32},\ref{eq:4nloh32},\ref{eq:4nlok32}),
we can define the Feynman rules for the perturbative calculation of the twist-2 pion wave function
$\phi^{(1),A}_{\pi}$  as a nonlocal hadronic matrix element with the structure
  $(\gamma^- \gamma_5)/2$ sandwiched:
\beq
\phi^{(1),A}_{\pi}=\frac{1}{2N_c P^-_2} \int \frac{dy^+}{2\pi} e^{-i x p^-_2 y^+}
  <\pi(p_2) | \overline{q}(y^+) (-i g_s) \int^{y^+}_{0} dz n \cdot A(zn)
\frac{\gamma^- \gamma_5}{2} q(0)| 0>,  \label{eq:nlopionA}
\eeq
which has the same form as the in the $\pi \gamma^\star \to \pi$\cite{prd64-014019}.
The relevant effective diagrams for the pion meson wave function are also showed in Fig.~\ref{fig:fig5},
and here only the first four diagrams in Fig.~\ref{fig:fig5} are useful to this sort NLO corrections
described in Fig.~\ref{fig:fig4} because the corrections with the gluon momentum partly flowing into LO hard kernel are cancelled in Eqs.~(\ref{eq:4nloup32},\ref{eq:4nlodown32}).
We can also derived the Feynman rule ($n_{\rho'}/(n\cdot l)$) for the Wilson line in Fig.~\ref{fig:fig5}
from the ${\calo}(\alpha_s)$ component of the pion wave function by the similar Fourier transformation
as for Fig.~\ref{fig:fig3}.
Then collinear factorization is therefore valid for the NLO corrections for the Fig.~\ref{fig:fig1}(a)
when  the additional gluon emitted from the final pion meson.

\subsection{${\cal O}(\alpha_s)$ correction to Fig.~\ref{fig:fig1}(b)}

\begin{figure}[tb]
\vspace{-1cm}
\begin{center}
\leftline{\epsfxsize=9cm\epsffile{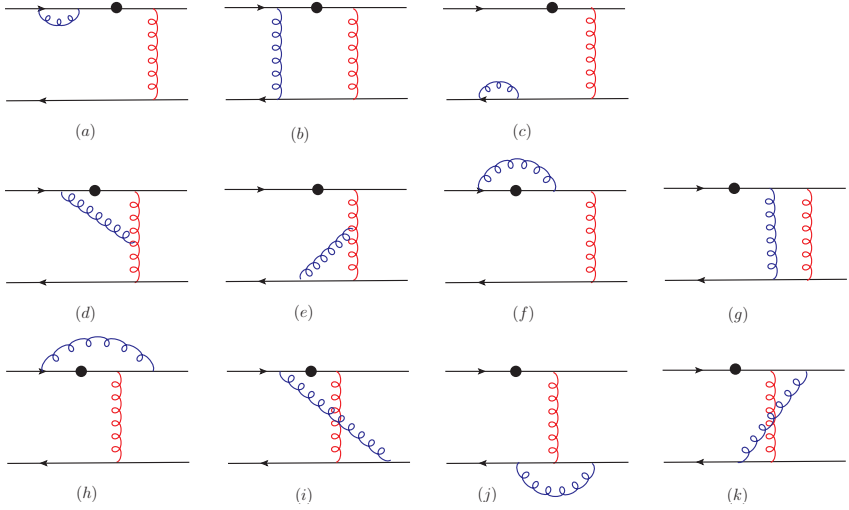}}
\end{center}
\vspace{-5.5cm}
\caption{$\calo(\alpha_s)$ correction to Fig.~\ref{fig:fig1}(b) with the additional gluon
(blue curves) emitted from the initial $\rho$ meson.}
\label{fig:fig6}
\end{figure}

\begin{figure}[tb]
\vspace{-1cm}
\begin{center}
\leftline{\epsfxsize=9cm\epsffile{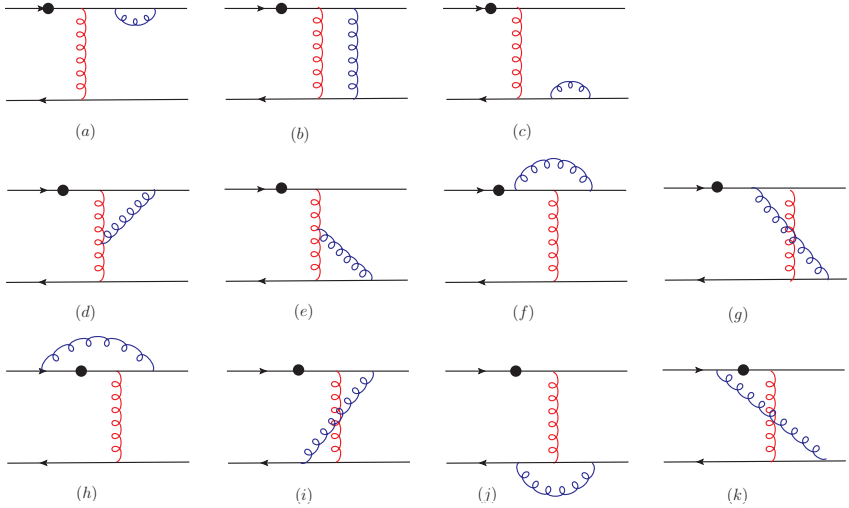}}
\end{center}
\vspace{-5.5cm}
\caption{$\calo(\alpha_s)$ corrections to Fig.~\ref{fig:fig1}(b) with the additional gluon
(blue curves) emitted from the final $\pi$ meson.}
\label{fig:fig7}
\end{figure}

In this subsection, we study the feasibility of the collinear factorization for the
NLO corrections to Fig.~\ref{fig:fig1}(b) with the same approach as we did for
Fig.~\ref{fig:fig1}(a).
With the requirement to hold the LO contents as shown in Eqs.~(\ref{eq:lob23},\ref{eq:lob32})
in the NLO factorization proof, we will consider both the T2\&T3 and T3\&T2 sets for
the DAs of the initial and final state meson in the NLO transition process
as illustrated in Fig.~\ref{fig:fig6} and Fig.~\ref{fig:fig7}.
We try to use the collinear factorization approach to separate the infrared
divergences of the amplitudes for Fig.~\ref{fig:fig6} and Fig.~\ref{fig:fig7} with the additional
blue gluons radiated from the initial rho meson and final pion meson respectively.


Firstly, the reducible sub-diagrams Figs.~\ref{fig:fig6}(a,b,c) and Figs.~\ref{fig:fig7}(a,b,c)
are factorized easily by simple inserting of the Fierz identity defied in Eq.~\ref{eq:fierz}.
And the soft divergences will be canceled exactly in these reducible amplitudes similarly as we
verified for Figs.~\ref{fig:fig2}(a,b,c)and Figs.~\ref{fig:fig4}(a,b,c).
We can then extract out the NLO twist-2 transversal rho meson DA's and the NLO twist-3 pion meson DA's in
the following forms:
\beq
\phi^{(1),T}_{\rho, a} &=& \frac{- i g^2_s C_F}{8} \frac{[\gamma^a_{\bot} \gamma^-] [\gamma_{\bot a} \gamma^+]
    (\psl_1-\ksl_1) \gamma^{\rho'} (\psl_1-\ksl_1+\lsl) \gamma_{\rho'}}{(p_1-k_1)^2 (p_1-k_1+l)^2 l^2}, \non
\phi^{(1),T}_{\rho, b} &=& \frac{i g^2_s C_F}{8} \frac{[\gamma^a_{\bot} \gamma^-] \gamma_{\rho'}
    (\ksl_1-\lsl) [\gamma_{\bot a} \gamma^+](\psl_1-\ksl_1+\lsl) \gamma^{\rho'}}{(k_1-l)^2 (p_1-k_1+l)^2 l^2}, \non
\phi^{(1),T}_{\rho, c} &=& \frac{- i g^2_s C_F}{8} \frac{[\gamma^a_{\bot} \gamma^-] \gamma^{\rho'}
    (\ksl_1-\lsl) \gamma_{\rho'} \ksl_1 [\gamma_{\bot a} \gamma^+]}{(k_1)^2 (k_1-l)^2 l^2};
\label{eq:6nloabcrhoT} \\
\phi^{(1),P}_{\pi, a} &=& \frac{- i g^2_s C_F}{4} \frac{\gamma_5 \gamma^{\rho'} (\psl_2-\ksl_2+\lsl) \gamma_{\rho'}
    (\psl_2-\ksl_2) \gamma_5}{(p_2-k_2)^2 (p_2-k_2+l)^2 l^2}, \non
\phi^{(1),P}_{\pi, b} &=& \frac{i g^2_s C_F}{4} \frac{(\ksl_2-\lsl) \gamma_{\rho'} \gamma_5
    \gamma^{\rho'} (\psl_2-\ksl_2+\lsl) \gamma_5}{(k_2-l)^2 (p_2-k_2+l)^2 l^2}, \non
\phi^{(1),P}_{\pi, c} &=& \frac{- i g^2_s C_F}{4}
\frac{\gamma_5 \ksl_2 \gamma^{\rho'} (\ksl_2-\lsl) \gamma_{\rho'} \gamma_5}
    {(k_1)^2 (k_1-l)^2 l^2}.
\label{eq:7nloabcpionP}
\eeq

Secondly, the NLO transversal NLO twist-2 rho meson DA $\phi^{(1),T}_{\rho, d}$ and the NLO twist-3 pion meson DAs
$\phi^{(1),P}_{\pi, d}$) can be extracted from the irreducible sub-diagrams Figs.~\ref{fig:fig6}(d,e,f,g) and
Figs.~\ref{fig:fig7}(d,e,f,g) respectively and are of the following form:
\beq
\phi^{(1),T}_{\rho, d} &=& \frac{- i g^2_s C_F}{8} \frac{[\gamma^a_{\bot} \gamma^-] [\gamma_{\bot a} \gamma^+]
    (\psl_1-\ksl_1+\lsl) \gamma^{\rho} v_{\rho}}{(p_1-k_1+l)^2 l^2 (v \cdot l)}.
\label{eq:6nlodrhoT} \\
\phi^{(1),P}_{\pi, d} &=& \frac{- i g_s^2 C_F}{4}
    \frac{\gamma_5 \gamma^{\rho} (\psl_2-\ksl_2+\lsl) \gamma_5 v_{\rho}}{(p_2-k_2+l)^2 l^2 (v \cdot l)},
\label{eq:7nlodpionP}
\eeq

Thirdly, the NLO transversal NLO rho meson DA $\phi^{(1),T/v/a}_{\rho, e}$ and the NLO twist-3 pion meson DAs
$\phi^{(1),P}_{\pi, e}$ can also be extracted from the
irreducible sub-diagrams Figs.~\ref{fig:fig6}(h,i,j,k) and Figs.~\ref{fig:fig7}(h,i,j,k) respectively,
and can be written in the following form:
\beq
\phi^{(1),T}_{\rho, e} &=& \frac{i g^2_s C_F}{8} \frac{[\gamma^a_{\bot} \gamma^-] \gamma^{\rho} (\ksl_1-\lsl)
    [\gamma_{\bot a} \gamma^+]  v_{ \rho}}{(k_1-l)^2 l^2 (v \cdot l)}
\left [1 - \frac{(k_1-k_2)^2}{(k_1-k_2-l)^2}\right ], \non
\phi^{(1),v}_{\rho, e} &=& \frac{i g^2_s C_F}{4} \frac{\gamma^b_{\bot} \gamma^{\rho} (\ksl_1-\lsl) \gamma_{a} v_{\rho}}
    {(k_1-l)^2 l^2 (v \cdot l)}
\left [1 - \frac{(k_1-k_2)^2}{(k_1-k_2-l)^2} \right ], \non
\phi^{(1),a}_{\rho, e} &=& \frac{i g^2_s C_F}{4} \frac{[\gamma_5 \gamma^{\mu'}_{\bot}] \gamma^{\rho} (\ksl_1-\lsl)
    [\gamma_{\bot \mu'} \gamma_5] v_{\rho}} {(k_1-l)^2 l^2 (v \cdot l)}
\left [1 - \frac{(k_1-k_2)^2}{(k_1-k_2-l)^2} \right ];
\label{eq:6nloerho}\\
\phi^{(1),P}_{\pi, e} &=& \frac{i g_s^2 C_F}{4} \frac{\gamma_5 (\ksl_2-\lsl) \gamma^{\rho} \gamma_5 n_{\rho}}
    {(k_2-l)^2 l^2 (n \cdot l)}.
\label{eq:7nloepionP}
\eeq

The hard LO amplitude $G^{(0)}_{b,23}(\xi_1, x_2), G^{(0),v/a}_{b,32}(\xi_1, x_2)$ and
$G^{(')(0)}_{b,23}(x_1,\xi_1, x_2)$ with the gluon momenta flowing or partly flowing into the original
LO hard amplitudes, are defined in the collinear region $l \parallel p_1$ for Fig.~\ref{fig:fig6} in the
following form:
\beq
&&G^{(0)}_{b,23}(\xi_1; x_2) =  \frac{i e g^2_s C_F}{2} \frac{[\epsl_{1T} \psl_1 \phi^{T}_{\rho}]
    \gamma^{\alpha} [\gamma_5 m^0_{\pi} \phi^P_{\pi}] \gamma_{\alpha} (\psl_2 - \ksl_1 +\lsl) \gamma_{\mu}}
    {(p_2-k_1+l)^2 (k_1-k_2-l)^2},
\label{eq:lob23i1} \\
&&G^{(0),v}_{b,32}(\xi_1; x_2) =  \frac{i e g^2_s C_F}{2} \frac{[\epsl_{1T} M_{\rho} \phi^{v}_{\rho}]
    \gamma^{\alpha} [\gamma_5 \psl_2 \phi^A_{\pi}] \gamma_{\alpha} (\psl_2 - \ksl_1 +\lsl) \gamma_{\mu}}
    {(p_2-k_1+l)^2 (k_1-k_2-l)^2},
\label{eq:lob32i1v} \\
&&G^{(0),a}_{b,32}(\xi_1; x_2) =  \frac{i e g^2_s C_F}{2}
    \frac{[M_{\rho} i \epsilon_{\mu'\nu\rho\sigma} \gamma_5 \gamma^{\mu'} \epsilon^{\nu}_{1T} n^{\rho} v^{\sigma}]
    \gamma^{\alpha} [\gamma_5 \psl_2 \phi^A_{\pi}] \gamma_{\alpha} (\psl_2 - \ksl_1 +\lsl) \gamma_{\mu}}
    {(p_2-k_1+l)^2 (k_1-k_2-l)^2},
\label{eq:lob32i1a} \\
&&G^{(0)}_{b,23}(x_1, \xi_1; x_2) =  \frac{i e g^2_s C_F}{2} \frac{[\epsl_{1T} \psl_1 \phi^{T}_{\rho}]
    \gamma^{\alpha} [\gamma_5 m^0_{\pi} \phi^P_{\pi}] \gamma_{\alpha} (\psl_2-\ksl_1+\lsl) \gamma_{\mu}}
    {(p_2-k_1+l)^2 (k_1-k_2)^2},
\label{eq:lob23i2} \\
&&G^{'(0)}_{b,23}(x_1, \xi_1; x_2) =  \frac{i e g^2_s C_F}{2} \frac{[\epsl_{1T} \psl_1 \phi^{T}_{\rho}]
    \gamma^{\alpha} [\gamma_5 m^0_{\pi} \phi^P_{\pi}] \gamma_{\alpha} (\psl_2-\ksl_1) \gamma_{\mu}}
    {(p_2-k_1)^2 (k_1-k_2-l)^2},
\label{eq:lob23i3}
\eeq
where the $\gamma^{\alpha}$ in Eqs.~(\ref{eq:lob23i2},\ref{eq:lob23i3}) could be
$\gamma^+$ or $\gamma^{\alpha}_{\bot}$.
When we set $\gamma^{\alpha}=\gamma^+$, the amplitude $G^{(0)}_{b,23}(x_1, \xi_1, x_2)$
becomes $G^{(0),L}_{b,23}(x_1, \xi_1, x_2)$, while $G^{'(0)}_{b,23}(x_1, \xi_1, x_2)$ becomes
$G^{'(0),L}_{b,23}(x_1, \xi_1, x_2)$.
When we choose $\gamma^{\alpha}=\gamma^{\alpha}_{\bot}$, the amplitude $G^{(0)}_{b,23}(x_1, \xi_1, x_2)$
becomes $G^{(0),T}_{b,23}(x_1, \xi_1, x_2)$, while $G^{'(0)}_{b,23}(x_1, \xi_1, x_2)$ becomes
$G^{'(0),T}_{b,23}(x_1, \xi_1, x_2)$.
And we can find that in the collinear region $l \parallel p_1$, these two newly defined LO
amplitudes in Eqs.~(\ref{eq:lob23i2},\ref{eq:lob23i2}) should be equal.

From the irreducible sub-diagrams of Fig.~\ref{fig:fig7} in the collinear region $l \parallel p_2$,
the LO hard amplitudes $G^{(0)}_{b,23/32}(x_1; \xi_2)$ and $G^{('')(0)}_{b,32}(x_1; \xi_2, x_2)$ with
the gluon momentum flowing or partly flowing into  the original LO hard amplitudes can be defined in the following form:
\beq
&&G^{(0)}_{b,23}(x_1; \xi_2) =  \frac{i e g^2_s C_F}{2} \frac{[\epsl_{1T} \psl_1 \phi^{T}_{\rho}]
    \gamma^{\alpha} [\gamma_5 m^0_{\pi} \phi^P_{\pi}] \gamma_{\alpha} (\psl_2-\ksl_1) \gamma_{\mu}}
    {(p_2-k_1)^2 (k_1-k_2-l)^2},
\label{eq:lob23f1} \\
&&G^{(0)}_{b,32}(x_1; \xi_2) =  \frac{i e g^2_s C_F}{2}
    \frac{[\epsl_{1T} M_{\rho} \phi^v_{\rho} + i M_{\rho} \epsilon_{\mu' \nu \rho \sigma} \gamma_5 \gamma^{\mu'}
    \epsilon^{\nu}_{1T} n^{\rho} v^{\sigma} \phi^a_{\rho}]} {(p_2-k_1)^2 (k_1-k_2-l)^2} \non
&&~~~~~~~~~~~~~~~~~~~~~~~ \cdot \gamma^{\alpha} [\gamma_5 \psl_2 \phi^A_{\pi}] \gamma_{\alpha} (\psl_2-\ksl_1) \gamma_{\mu},
\label{eq:lob32f1} \\
&&G^{(0)}_{b,32}(x_1; \xi_2, x_2) = \frac{i e g^2_s C_F}{2}
    \frac{[\epsl_{1T} M_{\rho} \phi^v_{\rho} + i M_{\rho} \epsilon_{\mu' \nu \rho \sigma} \gamma_5 \gamma^{\mu'}
    \epsilon^{\nu}_{1T} n^{\rho} v^{\sigma} \phi^a_{\rho}]} {(p_2-k_1+l)^2 (k_1-k_2)^2} \non
&&~~~~~~~~~~~~~~~~~~~~~~~ \cdot \gamma^{\alpha} [\gamma_5 \psl_2 \phi^A_{\pi}] \gamma_{\alpha} (\psl_2-\ksl_1) \gamma_{\mu},
\label{eq:lob32f2} \\
&&G^{''(0)}_{b,32}(x_1; \xi_2, x_2) = \frac{i e g^2_s C_F}{2}
    \frac{[\epsl_{1T} M_{\rho} \phi^v_{\rho} + i M_{\rho} \epsilon_{\mu' \nu \rho \sigma} \gamma_5 \gamma^{\mu'}
    \epsilon^{\nu}_{1T} n^{\rho} v^{\sigma} \phi^a_{\rho}]} {(p_2-k_1+l)^2 (k_1-k_2+l)^2} \non
&&~~~~~~~~~~~~~~~~~~~~~~~ \cdot \gamma^{\alpha} [\gamma_5 \psl_2 \phi^A_{\pi}] \gamma_{\alpha} (\psl_2-\ksl_1) \gamma_{\mu}.
\label{eq:lob32f3}
\eeq

By summing up the amplitudes from those irreducible sub-diagrams Figs.~\ref{fig:fig6}(d,e,f,g,h,i,j,k),
the total NLO amplitudes with the crossed-twist DAs T2\&T3 (i.e. the NLO set-I amplitude)
can be written in a convolution
of the NLO rho wave function and the LO hard amplitudes, with the gluon momentum not flowing, flowing or
partly flowing into the LO hard kernels:
\beq
G^{(1)}_{\rm 6,23}(x_1; x_2) &=& \phi^{(1),T}_{\rho,d} \otimes
  \Bigl \{G^{(0),L}_{b,23}(x_1; x_2) - \frac{9}{16} G^{(0),L}_{b,23}(\xi_1; x_2)
                                     - \frac{9}{16} G^{(0),L}_{b,23}(x_1,\xi_1; x_2) \non
   && ~~~~~~~~~~~                    + G^{(0),T}_{b,23}(x_1; x_2) - G^{(0),T}_{b,23}(\xi_1; x_2)
                                     + \frac{1}{8} G^{(0),T}_{b,23}(x_1, \xi_1; x_2)\Bigr \} \non
  +&& \phi^{(1),T}_{\rho,e} \otimes
  \Bigl \{- G^{(0),L}_{b,23}(\xi_1; x_2) + \frac{9}{16} G^{(0),L}_{b,23}(x_1; x_2)
                                    + \frac{9}{16} G^{(0),L}_{b,23}(x_1, \xi_1; x_2) \non
   && ~~~~~~~~~~~                   + G^{(0),T}_{b,23}(x_1; x_2) - G^{(0),T}_{b,23}(\xi_1; x_2)
                                    -\frac{1}{8} G^{(0),T}_{b,23}(x_1, \xi_1; x_2)\Bigr \}.
\label{eq:6nlo23}
\eeq
It's easy to find that the soft divergences from the collinear region for these irreducible amplitudes are
canceled each other. At the quark level, consequently, there is no soft divergence left after the
summation for the contributions from all sub-diagrams in Fig.~\ref{fig:fig6} with the case of the T2\&T3 DAs.
The collinear divergences, generated from the gluon radiated from the up-line quark and the down-line
anti-quark of the initial rho meson in Fig.~\ref{fig:fig6}, can be absorbed into the NLO twist-2 rho meson
DA $\phi^{(1),T}_{\rho,d/e}$, which is written in the following nonlocal hadronic matrix element in
$\mathbf{b}$ space with the structure $(\gamma^b_{\bot} \gamma^+)/4$ sandwiched:
\beq
\phi^{(1),T}_{\rho}=\frac{1}{2N_c P^+_1} \int \frac{dy^-}{2\pi} e^{-i x p^+_1 y^-}
     <0 | \overline{q}(y^-) \frac{\gamma^b_{\bot} \gamma^+}{4} (-i g_s)
     \int^{y^-}_{0} dz v \cdot A(zv) q(0)| \rho(p_1)>.
\label{eq:nlorhoT}
\eeq

We then give a short summary for the NLO Set-II amplitudes with the crossed-twist DAs T3\&T2
for the sub-diagrams Figs.~\ref{fig:fig6}(d,e,f,g,h,i,j,k).
After the summation, the total infrared divergences from the NLO corrections to the LO hard amplitude
$G^{(0)}_{b,32}(x_1; x_2)$ in the $l \parallel p_1$ region can be written in the following form:
\beq
G^{(1)}_{6,32}(x_1; x_2) = \phi^{(1),v}_{\rho,e} \otimes \left \{2G^{(0),v}_{b,23}(x_1; x_2)
                                          + 2G^{(0),v}_{b,23}(\xi_1; x_2)\right \}.
\label{eq:6nlo32}
\eeq
We find that the infrared divergences from the Set-II amplitudes for sub-diagrams Figs.~\ref{fig:fig6}(d,g,h,i),
with the additional gluon radiated from the right-up quark line, are suppressed by the kinetic constraints,
then only the sub-diagrams Figs.~\ref{fig:fig6}(e,g,j,k) generate infrared divergent corrections to the Set-II
LO amplitudes $G^{(0)}_{b,32}(x_1; x_2)$ with T3\&T2 DAs.
The soft divergences from the sub-diagrams with the gluon radiated from the left-down anti-quark line were
canceled exactly.
Only the collinear divergences, generated from the gluon radiated from the left-down anti-quark of the
initial rho meson in Fig.~\ref{fig:fig6}, will be absorbed into the NLO twist-3 rho meson DA $\phi^{(1),v}_{\rho}$.
Then we can  factorize the Set-II irreducible amplitudes for Fig.~\ref{fig:fig6} in the collinear region
as the convolutions of the NLO twist-3 DA and LO hard amplitudes.
The collinear factorization is therefore valid for the NLO Set-II corrections for the Fig.~\ref{fig:fig1}(b)
with the additional gluon emitted from the initial rho meson.

Now, we elaborate the factorization for the infrared divergences in the irreducible sub-diagrams
Fig.~\ref{fig:fig7}(d,e,f,g,h,i,j,k), in which the additional blue gluons are radiated from the final pion meson.
The total NLO corrections for the Set-I amplitudes of the sub-diagrams Figs.~\ref{fig:fig7}(d,e,f,g,h,i,j,k)
with the T2\&T3 DAs from the $l \parallel p_2$ region are also summed over and can be written in the following
convolution form:
\beq
G^{(1)}_{\rm 7,23}(x_1; x_2) &=& \phi^{(1),P}_{\pi,d} \otimes
       \Bigl \{ \frac{7}{16}\left [G^{(0),L}_{b,23}(x_1; x_2) - G^{(0),L}_{b,23}(x_1; \xi_2)\right ]\non
&&               -\frac{9}{16}\left [G^{(0),T}_{b,23}(x_1; x_2) - G^{(0),T}_{b,23}(x_1; \xi_2)\right ] \Bigr \} \non
&&     + \phi^{(1),P}_{\pi,e} \otimes
            \frac{25}{16} \left[ G^{(0),T}_{b,23}(x_1; x_2)
                                    - G^{(0),T}_{b,23}(x_1; \xi_2) \right].
  \label{eq:7nlo23}
  \eeq
For the NLO Set-I amplitudes in Eq.~(\ref{eq:7nlo23}), the soft divergences from sub-diagrams
Figs.~\ref{fig:fig7}(h,i) can be canceled by their counterparts from Figs.~\ref{fig:fig7}(j,k),
then only the collinear divergences are left for the infrared absorbtion.
The collinear divergences can all be absorbed into the NLO pion meson DAs of $\phi^{(1),P}_{\pi,d/e}$,
which can be written as the nonlocal hadronic matrix element with the structure as that in Ref.~\cite{epjc40-395}:
\beq
\phi^{(1),P}_{\pi}=&&\frac{1}{2N_c P^-_2} \int \frac{dy^+}{2\pi} e^{-i x p^-_2 y^+}
    <\pi(p_2) | \overline{q}(y^+) (-i g_s) \int^{y^+}_{0} dz n \cdot A(zn) \frac{\gamma_5}{2} q(0)| 0>.
\label{eq:nlopionP}
\eeq
As shown in Eq.~(\ref{eq:7nlo23}), all Set-I infrared-relevant NLO amplitudes can be written
as the convolution of the LO hard kernel and the NLO $\pi$ meson DAs
($G^{(0)}_{b,23} \otimes \phi^{(1),P}_{\pi,d}$ and $G^{(0)}_{b,23} \otimes \phi^{(1),P}_{\pi,e}$),
with the integral momenta flowing or not flowing into the LO hard amplitudes.

By making the summation for the Set-II amplitudes for the sub-diagrams Figs.~\ref{fig:fig7}(d,e,f,g,h,i,j,k)
with the T3\&T2 DAs, we find:
\beq
G^{(1)}_{\rm 7,32}(x_1; x_2) &=& \phi^{(1),A}_{\pi,d} \otimes
        \Bigl \{ G^{(0)}_{b,32}(x_1; x_2) - G^{(0)}_{b,32}(x_1; \xi_2)
           +\frac{1}{8} G^{(0)}_{b,32}(x_1; \xi_2, x_2) \Bigr \} \non
&& + \phi^{(1),A}_{\pi,e} \otimes \Bigl \{G^{(0)}_{b,32}(x_1; x_2) - G^{(0)}_{b,32}(x_1; \xi_2)
                                   -\frac{1}{8} G^{''(0)}_{b,32}(x_1; \xi_2, x_2) \Bigr\}.
\label{eq:7nlo32}
\eeq
For the infrared singularities in the Set-II amplitudes in Eq.~(\ref{eq:7nlo32}), the soft divergences
from Figs.~\ref{fig:fig7}(i,j) can't be canceled by their counterparts in Figs.~\ref{fig:fig7}(h,k),
but these soft divergences are also diminished because $\phi^{(1),A}_{\pi,e}=\phi^{(1),A}_{\pi,d}$.
The remaining collinear singularities in Eq.~(\ref{eq:7nlo32}) can be absorbed into the NLO DAs
$\phi^{(1),A}_{\pi,d/e}$.
All these irreducible NLO amplitudes can be written as the convolution of the LO hard kernel
and the NLO $\pi$ meson DAs($G^{(0)}_{b,23} \otimes \phi^{(1),A}_{\pi,e}$
and $G^{(0)}_{b,32} \otimes \phi^{(1),A}_{\pi,e}$),  and the collinear factorization
approach is valid for the Fig.~\ref{fig:fig7}.


\section{$k_T$ factorization Of $\rho \gamma^{\star} \to \pi$}

In this section, the NLO proof of the factorization theorem is demonstrated with the inclusion
of the transversal momentum $k_T$.
The $k_T$ factorization approach is qualified to deal with the
small-x physics\cite{plb87-359,prd64-014019,zpz50-139},
because of it's advantage to avoid the end-point singularity without introducing
other non-physics methods.

\begin{figure}[tb]
\vspace{-1cm}
\begin{center}
\leftline{\epsfxsize=10cm\epsffile{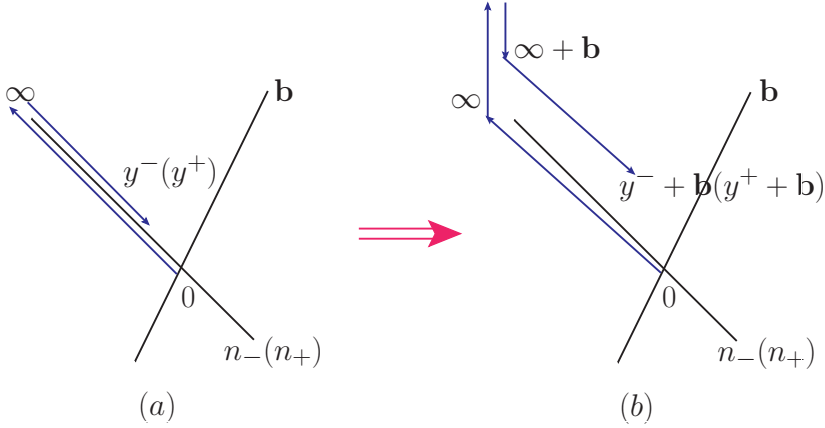}}
\end{center}
\vspace{-8.5cm}
\caption{The deviation of the integral (Wilson link) from the light corn by $\mathbf{b}$ in
the coordinate space for the two-parton meson wave function.}
\label{fig:fig8}
\end{figure}

The hierarchy $k^2_{iT} \ll k_1\cdot k_2$ is holding in the bound wave functions,
so the transversal contributions on the numerators can be dropped
safely and the transversal momentum $k_T$ in the LO hard kernels can also be dropped,
then factorization proofs made in the above section with the collinear factorization approach
is valid here with the inclusion of the transversal momentum \cite{prd67-034001,prd83-054029}.
When we extend the proofs for the NLO $\rho \to \pi$ transition from collinear factorization approach
to $k_T$ factorization approach,
the only modification required  is to include the transversal integral $l_T$ to
the NLO wave functions in Eqs.~(\ref{eq:nlorhovd},\ref{eq:nlorhoad},
\ref{eq:nlopionA},\ref{eq:nlorhoT},\ref{eq:nlopionP}),
besides the longitudinal integral along the light cone.
This modification can also be understood as the integral deviated from the light cone direction
by $\mathbf{b}$ in the coordinate space, as illustrated by Fig.~\ref{fig:fig8}.

The $\calo(\alpha_s)$ wave functions at twist-2 and twist-3  as deifined
in  Eqs.~(\ref{eq:nlorhovd},\ref{eq:nlorhoad},\ref{eq:nlopionA},\ref{eq:nlorhoT},\ref{eq:nlopionP})
can be reproduced by the following nonlocal matrix element in the $\mathbf{b}$ space.
\beq
  \phi^{(1),T}_{\rho}(x_1,\xi_1;\mathbf{b_1})=&&\frac{1}{2N_c P^+_1} \int \frac{dy^-}{2\pi} \frac{d\mathbf{b_1}}{(2\pi)^2}
  e^{-i x p^+_1 y^+ + i \mathbf{k_{1T}} \cdot \mathbf{b_1}} \non
  &&\cdot <0| \overline{q}(y^-) \frac{\gamma^b_{\bot} \gamma^+}{4} (-i g_s) \int^{y}_{0} dz v \cdot A(zv) q(0)|\rho(p_1)>,
  \label{eq:nlorhoTkt}
  \eeq
  \beq
  \phi^{(1),v}_{\rho}(x_1,\xi_1;\mathbf{b_1})=&&\frac{1}{2N_c P^+_1} \int \frac{dy^-}{2\pi} \frac{d\mathbf{b_1}}{(2\pi)^2}
  e^{-i x p^+_1 y^+ + i \mathbf{k_{1T}} \cdot \mathbf{b_1}} \non
  &&\cdot <0|\overline{q}(y^-) \frac{\gamma_{\bot}}{2} (-i g_s)  \int^{y}_{0} dz v \cdot A(zv) q(0)|\rho(p_1)>,
  \label{eq:nlorhovdkt}
  \eeq
  \beq
  \phi^{(1),a}_{\rho}(x_1,\xi_1;\mathbf{b_1})=&&\frac{1}{2N_c P^+_1} \int \frac{dy^-}{2\pi} \frac{d\mathbf{b_1}}{(2\pi)^2}
  e^{-i x p^+_1 y^+ + i \mathbf{k_{1T}} \cdot \mathbf{b_1}} \non
  &&\cdot <0|\overline{q}(y^-) \frac{\gamma_5 \gamma_{\bot}}{2} (-i g_s) \int^{y}_{0} dz v \cdot A(zv) q(0)|\rho(p_1)>;
  \label{eq:nlorhoadkt}
  \eeq
  \beq
  \phi^{(1),A}_{\pi}(\xi_2,x_2;\mathbf{b_2})=&&\frac{1}{2N_c P^-_2} \int \frac{dy^+}{2\pi} \frac{d\mathbf{b_2}}{(2\pi)^2}
  e^{-i x p^-_2 y^+ + i \mathbf{k_{2T}} \cdot \mathbf{b_2}} \non
  &&\cdot <\pi(p_2)|\overline{q}(y^+) (-i g_s) \int^{y}_{0} dz n \cdot A(zn) \frac{\gamma^- \gamma_5}{2} q(0)|0>,
  \label{eq:nlopionAkt}
  \eeq
  \beq
  \phi^{(1),P}_{\pi}(\xi_2,x_2;\mathbf{b_2})=&&\frac{1}{2N_c P^-_2} \int \frac{dy^+}{2\pi} \frac{d\mathbf{b_2}}{(2\pi)^2}
  e^{-i x p^-_2 y^+ + i \mathbf{k_{2T}} \cdot \mathbf{b_2}} \non
  &&\cdot <\pi(p_2)|\overline{q}(y^+) (-i g_s) \int^{y}_{0} dz n \cdot A(zn) \frac{\gamma_5}{2} q(0)|0>.
  \label{eq:nlopionPkt}
  \eeq
All these NLO wave functions would reproduce the Feynman rules of Wilson lines.

\section{Summarey}

In this paper we firstly verified that the factorization hypothesis is valid for the $\rho \to \pi$
transition process at NLO level in the collinear factorization approach, and then we extended
this proof to the case of the $k_T$ factorization approach.
Because of the difference of the initial vector meson $\rho$ and the final pseudo-scalar
meson $\pi$, we considered both the two LO sub-diagrams Figs.~\ref{fig:fig1}(a) and 1(b),
with the virtual photon vertex positioned on the initial state quark line and on the final state
quark line, respectively.

For each LO sub-diagram Fig.~\ref{fig:fig1}(a) or Fig.~\ref{fig:fig1}(b),
we first evaluated  the NLO corrections from the additional gluon radiated from
the initial rho meson as well as from the final pion meson,
and then we verified that all the infrared singularities in those four NLO quark level diagrams
( Fig.1(a) - Fig.1(d) ) could be absorbed into the NLO meson wave functions.
Certainly, we made this proof both in the collinear factorization approach
and in the $k_T$ factorization approach.
And we showed explicitly that every NLO quark level amplitude can be expressed
as the convolution of the NLO wave functions and the LO hard kernel, with the gluon momenta,
which would generate the infrared singularities, flowing, not flowing or partly flowing into
the LO hard amplitudes.

Particularly, we find that: (a) only the T3\&T2  set with the twist-3 $\rho$ meson DAs and
twist-2 pion DAs contribute to the LO amplitude of Fig.~1(a),  as defined in Eq.~(\ref{eq:loa32});
(b) only the collinear singularities would appeare in the NLO diagrams for the LO Fig.~\ref{fig:fig1}(a),
because the soft singularities in these NLO diagrams are either suppressed
by the kinetics or canceled each other.

For the NLO corrections to the LO Fig.~\ref{fig:fig1}(b), however,
there exist two kinds of the  LO amplitudes as described in Eqs.~(\ref{eq:lob23},\ref{eq:lob32})
with the T2\&T3 and T3\&T2 combinations of the initial and final state
meson wave functions and we called them Set-I and Set-II respectively.
We further find that the NLO corrections to the Set-I and Set-II LO amplitude
generate the collinear singularities  only, since the soft singularities in
these two cases are either suppressed by the kinetics or canceled each other.
The underlying reason is the fact that
the soft gluon will not change the color structure of the rho and pion mesons.
All the remaining infrared singularities from the collinear regions, should be absorbed into
the NLO wave functions,
and we have also defined the NLO wave functions with different twists in
the nonlocal matrix elements,
which would help us to understand the fundamental meson wave functions
and push us to calculate the NLO hard kernels for this $\rho \to \pi$ transition process.

\section{Acknowledement}

The authors would like to thank H.N.~Li and C.D.~Lu for long term collaborations
and valuable discussions.
This work is supported by the National Natural Science Foundation of China
under Grant No. 11235005,
and by the Project on Graduate Students¡¯ Education and Innovation of Jiangsu
Province, under Grant No. CXZZ13-0391.


\end{document}